\documentclass[11pt,a4paper]{article}
\pdfoutput=1
\usepackage{jheppub}

\usepackage{placeins}
\usepackage{epsfig}
\usepackage{latexsym}
\usepackage[bbgreekl]{mathbbol}
\usepackage{mathtools}
\usepackage{amsmath,amssymb,amsthm,amsbsy,amsfonts}
\usepackage{multirow}
\usepackage{slashed}
\usepackage{float}
\usepackage{esint}
\usepackage{epsfig}
\usepackage{lscape}
\usepackage{graphicx}
\usepackage{psfrag}
\usepackage[utf8]{inputenc}
\usepackage{subfigure}
\usepackage{nicefrac}
\usepackage[stable]{footmisc}
\usepackage{rotating}
\usepackage{afterpage}
\usepackage{bbm}
\usepackage{color,xcolor}
\usepackage{cancel}

\def\){\right)}
\def\({\left( }
\def\]{\right] }
\def\[{\left[ }

\def\NO{\nonumber}

\newcommand{\be}{\begin{equation}}
\newcommand{\ee}{\end{equation}}

\def\bea{\begin{eqnarray}}
\def\eea{\end{eqnarray}}

\def\bal#1\eal{\begin{align}#1\end{align}}

\def\bald{\begin{aligned}}
\def\eald{\end{aligned}}

\def\bsub{\begin{subequations}}
\def\esub{\end{subequations}}

\def\beqx{\begin{displaymath}}
\def\eeqx{\end{displaymath}}

\newcommand{\bmat}{\left(\begin{array}}
\newcommand{\emat}{\end{array}\right)}

\def\a{\alpha}
\def\b{\beta}
\def\c{\chi}
\def\d{\delta}
\def\e{\epsilon}
\def\f{\phi}
\def\g{\gamma}

\def\k{\kappa}
\def\l{\lambda}
\def\m{\mu}
\def\n{\nu}
\def\o{\omega}
    
\def\p{\pi}

    \def\th{\theta}
\def\r{\rho}
\def\s{\sigma}
\def\t{\tau}
\def\x{\xi}

\def\D{\Delta}
\def\F{\Phi}
\def\G{\Gamma}
\def\J{\Psi}
\def\L{\Lambda}
\def\O{\Omega}
    
\def\P{\Pi}

\def\S{\Sigma}

\def\ve{\varepsilon}

\def\vf{\varphi}

\def\cf{{\cal F}}
\def\cg{{\cal G}}
\def\ch{{\cal H}}
\def\ci{{\cal I}}
\def\cj{{\cal J}}
\def\ck{{\cal K}}

\def\cm{{\cal M}}
\def\cn{{\cal N}}
\def\co{{\cal O}}

\def\car{{\cal R}}
\def\cs{{\cal S}}
\def\ct{{\cal T}}
\def\cu{{\cal U}}
\def\cv{{\cal V}}
\def\cw{{\cal W}}
\def\cx{{\cal X}}
\def\cy{{\cal Y}}

\def\bb#1{\ensuremath{\mathbb{#1}}} 

\def\bo{{\raise-.3ex\hbox{\large$\Box$}}}               
\def\pa{\partial}                                       
\def\face{{\raise.2ex\hbox{$\displaystyle \bigodot$}\mskip-2.2mu \llap {$\ddot
        \smile$}}}                                   
\def\>{\rangle}                                      
\def\<{\langle}                                      

\def\tx#1{\text{#1}}
\def\sbtx#1{{}_{\rm #1}}                           
\def\lbar#1{\ensuremath{\overline{#1}}}              
\def\leftrightarrowfill{$\mathsurround=0pt \mathord\leftarrow \mkern-6mu
        \cleaders\hbox{$\mkern-2mu \mathord- \mkern-2mu$}\hfill
        \mkern-6mu \mathord\rightarrow$}        
\def\dvec#1{\vbox{\ialign{##\crcr
        \leftrightarrowfill\crcr\noalign{\kern-1pt\nointerlineskip}
        $\hfil\displaystyle{#1}\hfil$\crcr}}}           
\def\Re{{\rm Re\,}}                                     
\def\Im{{\rm Im\,}}                                     

\def\-{\hphantom{-}}

\title{Entropy functional and the holographic attractor mechanism}

\author[a]{Alejandro Cabo-Bizet}
\author[b,c]{Uri Kol}
\author[c,d]{Leopoldo A. Pando Zayas}
\author[e]{Ioannis Papadimitriou} 
\author[c]{Vimal Rathee}

\affiliation[a]{Department of Mathematics, King’s College London, The Strand, London WC2R 2LS, UK}
\affiliation[b]{Center for Cosmology and Particle Physics, Department of Physics, New York University, 726 Broadway, New York, NY 10003, USA}
\affiliation[c]{Leinweber Center for Theoretical Physics, University of Michigan, Ann Arbor, MI 48109, USA}
\affiliation[d]{The Abdus Salam International Centre for Theoretical Physics, Strada Costiera 11, 34014 Trieste, Italy}
\affiliation[e]{School of Physics, Korea Institute for Advanced Study, Seoul 02455, Korea}

\emailAdd{acbizet@gmail.com}
\emailAdd{urikol@nyu.edu}
\emailAdd{lpandoz@umich.edu}
\emailAdd{ioannis@kias.re.kr}
\emailAdd{vimalr@umich.edu}

\abstract{We provide a field theory interpretation of the attractor mechanism for asymptotically AdS$_4$ dyonic BPS black holes whose entropy is captured by the supersymmetric index of the twisted ABJM theory at Chern-Simons level one. We holographically compute the renormalized off-shell quantum effective action in the twisted ABJM theory as a function of the supersymmetric fermion masses and the arbitrary vacuum expectation values of the dimension one scalar bilinear operators and show that extremizing the effective action with respect to the vacuum expectation values of the dimension one scalar bilinears is equivalent to the attractor mechanism in the bulk. In fact, we show that the holographic quantum effective action coincides with the entropy functional and, therefore, its value at the extremum reproduces the black hole entropy. 
}


\preprint{KIAS-P17130, LCTP-17-07}

\begin{document}  
	\maketitle



\section{Introduction}
\label{intro}

The AdS/CFT correspondence conjectures a mathematical equivalence between string theories, containing gravity, and field theories \cite{Maldacena:1997re}. This powerful paradigm is capable of translating puzzling black hole physics into questions in a unitary field theory. This correspondence provides, in principle, key answers regarding not only the nature of the microscopic degrees of freedom responsible for the macroscopic entropy of black holes, but also the resolution of the information paradox. 
	
Only recently, however, has an explicit realization of AdS$_4$/CFT$_3$  yielded impressive results for the microstate counting of black hole entropy \cite{Benini:2015eyy}.  Under a series of assumptions, more crucially an identification of chemical potentials and an extremization procedure, a perfect large $N$ match between the topologically twisted index and the black hole entropy was established  \cite{Benini:2015eyy}.   Under similar assumptions  matches have now been established in various other situations including: dyonic black holes \cite{Benini:2016rke}, black holes with hyperbolic horizons   \cite{Cabo-Bizet:2017jsl}, and black holes in massive IIA theory  \cite{Azzurli:2017kxo,Benini:2017oxt, Hosseini:2017fjo}.  
	
Our goal is largely motivated by a desire to conceptually clarify, within the standard AdS/CFT dictionary, the various assumptions made in \cite{Benini:2015eyy}. Consider, for example, the role of the attractor mechanism which is a key intuition building concept in our understanding of black holes in supergravity theories \cite{Ferrara:1995ih,Strominger:1996kf,Ferrara:1996dd}. It roughly states that the black hole entropy is determined by  extremization of the central charge in the moduli space.  A decade after its original formulation, the attractor mechanism intuition took an upgraded incarnation - the entropy formula \cite{Sen:2005wa} - which accommodates higher curvature corrections and weakens the hold of supersymmetry.  There is, however, an important conceptual difference between the attractor mechanism in flat  space and its counterpart in asymptotically AdS spacetimes. In asymptotically flat spacetimes the attractor mechanism is loosely associated with no-hair theorems. In asymptotically AdS spacetimes this intuition is lacking due to the natural existence of hair. Moreover, in the context of the AdS/CFT most of the key properties of the duality are precisely defined in the asymptotic region, not close to the horizon. This dichotomy between boundary and horizon data has been pointed out before and discussed in the context of  Wald entropy formula in \cite{Dutta:2006vs}. Here we address it via the AdS/CFT correspondence.

Recall that in the attractor mechanism one extremizes the central charge with respect to the moduli \cite{Ferrara:1996dd}. We demonstrate that for asymptotically AdS black holes in gauged supergravity, the attractor mechanism can be reinterpreted using exclusively boundary data. More precisely, using the AdS/CFT dictionary, we compute the renormalized off-shell quantum effective action in the twisted ABJM theory as a function of the supersymmetric fermion masses and the arbitrary vacuum expectation values of the dimension one scalar bilinear operators. This effective action coincides with the entropy functional and we show that its extremization with respect to the vacuum expectation values of the dimension one scalar bilinears is equivalent to the attractor mechanism. We thus provide a strictly field theoretic interpretation of the attractor mechanism in the context of ${\cal N}=2$ gauged supergravity and a rigorous understanding of the beautiful results of \cite{Benini:2015eyy}. 
	
The manuscript is organized as follows. In section \ref{Sec:N2}, we review the relevant structure of  ${\cal N}=2$ gauge supergravity and provide a universal formula for the regularized on-shell action in terms of the effective superpotential for general dyonic black holes introduced in \cite{Lindgren:2015lia}. Section \ref{holren} is devoted to key aspects of the  holographic dictionary. We derive the supersymmetric boundary counterterms and discuss the supersymmetric boundary conditions for the scalars. Moreover, we determine the renormalized operators dual to bulk fields and we compute the renormalized quantum effective action for dyonic BPS black holes. Using this quantum effective action, in section \ref{attractor} we obtain one of the key results of the paper: a holographic interpretation of the attractor mechanism. We conclude in section \ref{conclusion}. Some technical details are relegated to two appendices. In appendix \ref{STU-parameterizations} we explicitly discuss various parameterizations of the STU model, and in \ref{ham} we review the radial Hamiltonian formulation of the bulk dynamics.

\section{Effective superpotential for dyonic black holes}
\label{Sec:N2}

We are mostly interested in black hole solutions of the Abelian $U(1)^4$ $\mathcal{N} = 2$ gauged supergravity in four dimensions, often referred to as the gauged STU model, which is a consistent truncation of $\mathcal{N} = 8$ gauged supergravity \cite{Duff:1999gh,Cvetic:1999xp}. With appropriate supersymmetric boundary conditions, this theory is holographically dual to a sector of the ABJM theory at Chern-Simons level one. Most of our analysis, however, applies broadly to ${\cal N}=2$ gauged supergravity and we begin by briefly reviewing some general properties.

\subsection{$\cn=2$ gauged supergravity}
\label{N=2}
As an example of the generality of our approach we describe the $U(1)^4$ theory using the general framework of $\cn = 2$ gauged supergravity in four dimensions. In this language the $U(1)^4$ theory consists of the gravity multiplet coupled to $n_V=3$ vector multiplets and no hypermultiplets. Since the gauge group
is Abelian, the scalars in the vector multiplets are neutral and so the only charged fields present are the two gravitini. This is usually referred to as Fayet-Iliopoulos (FI) gauging.\footnote{Throughout this paper we consider only purely electric gauging.} The
gauge fields that couple to the gravitini are a linear combination of the graviphoton and
the $n_V$ vectors from the vector multiplets, $\x_\L A^\L_\m$ , with $\L = 0, 1, \ldots , n_V$. The constants $\x_\L$ are called the FI parameters. For the $U(1)^4$ theory the FI parameters are all equal, i.e. 
\be\label{FI-parameters}
\x_0=\x_1=\x_2=\x_3=\x>0,
\ee 
where the value of the constant $\x$ depends on the normalization of the vector fields in the Lagrangian. For general FI parameters we define $2\x\equiv\sqrt{\x_0^2+\x_1^2+\x_2^2+\x_3^2}$. We keep $\x$ arbitrary in order to facilitate comparison with different conventions in the literature. 

The complex scalars $z^\a$ in the vector multiplets, with $\a = 1, \ldots , n_V$, parameterize a special Kähler manifold -- an $n_V$-dimensional Hodge-Kähler manifold which is the base of a symplectic bundle with the covariantly holomorphic sections
\be
\frak V= e^{\ck(z,\bar z)/2} (X^\L , F_\L),
\ee
where $\ck$ is the K\" ahler potential. In certain symplectic frames there exist a second degree homogeneous function $F(X)$, called the prepotential, such that $F_\L = \pa_\L F$. For the STU model (in the duality frame of purely electric gaugings) the prepotential is 
\be\label{prepotential}
F=-2i\sqrt{X^0X^1X^2X^3},
\ee
and so
\be
F_\L=\frac{F}{2X^\L}.
\ee
The holomorphic sections define the embedding ambient space
\be
\<\frak V, \frak V\>\equiv e^{\ck(z,\bar z)} \(X^\L \lbar F_\L-\lbar X^\L F_\L\) = i,
\ee
which in turn defines the Kähler potential in terms of the holomorphic sections
\be\label{Kahler-potential}
\ck=-\log\Big(i(\lbar X^\L F_\L-X^\L\lbar F_\L)\Big).
\ee
The corresponding K\"ahler metric is given by
\be\label{Kahler-metric}
\ck_{\a\bar\b}=\pa_\a\pa_{\bar\b}\ck.
\ee

The above data completely determines the bosonic part of the $\cn=2$ gauged  supergravity action to be 
\be\label{action}
S=\frac{1}{2\k^2}\int_\cm d^{4}x\sqrt{-g}\Big(R-\cg_{\a\bar\b}\pa^\m z^\a\pa_\m \bar z^{\bar\b} -2\ci_{\L\S}F^\L_{\m\n}F^{\S\m\n}-\car_{\L\S}\e^{\m\n\r\s}F^\L_{\m\n}F^{\S}_{\r\s}-\cv\Big)+S\sbtx{GH},
\ee
where  
\be\label{GH}
S\sbtx{GH}=\frac{1}{2\k^2}\int_{\pa\cm} d^3x \sqrt{-\g}\;2K,
\ee
is the Gibbons-Hawking term and we have normalized the fields such that the gravitational constant $\k^2=8\p G_{4}$ appears as an overall factor in front of the action, as appropriate for comparing our results with the large-$N$ limit of the dual ABJM theory. We use the standard $-\; +\; +\; +$ signature for the metric and we have reversed the signs of the symmetric matrices $\ci_{\L\S}$ and $\car_{\L\S}$ relative to the usual convention in the supergravity literature (e.g. \cite{DallAgata:2010ejj,Klemm:2016wng}) since with our definition the eigenvalues of $\ci_{\L\S}$ are positive definite. Moreover, with our normalization of the vector multiplet scalars the scalar metric is related to the K\"ahler metric \eqref{Kahler-metric} as 
\be\label{scalar-metric}
\cg_{\a\bar\b}=2\ck_{\a\bar\b}=2\pa_\a\pa_{\bar\b}\ck.
\ee

The real symmetric matrices $\ci_{\L\S}$ and $\car_{\L\S}$ are given by
\be
\ci_{\L\S}=-\Im\cn_{\L\S},\qquad \car_{\L\S}=-\Re\cn_{\L\S},\qquad \det(\ci) >0,
\ee
where the period matrix $\cn_{\L\S}$ is defined through the relations 
\be
F_\L=\cn_{\L\S}X^\S,\qquad \pa_{\bar \a}\lbar F_\L=\cn_{\L\S}\pa_{\bar\a}\lbar X^\S.
\ee
Whenever a prepotential exits the period matrix can be expressed as (see e.g. \cite{Klemm:2016wng}) 
\be
\cn_{\L\S}=\lbar F_{\L\S}+2 i\frac{\Im(F_{\L P})X^P \Im(F_{\S\F})X^\F}{X^\O \Im(F_{\O\J})X^\J},
\ee
where 
\be
F_{\L\S}\equiv \pa_\L F_\S=\pa_\L\pa_\S F=\frac{F}{4X^\L X^\S}\(1-2\d^{\L\S}\).
\ee
The last equality applies only to the STU model prepotential \eqref{prepotential}.

Finally, the scalar potential is obtained from the holomorphic superpotential
\be\label{hol-sup}
W\equiv\sum_\L \x_\L X^\L,
\ee
through the identity
\be
\x^2 L^2\cv=e^{\ck}\(\ck^{\a\bar\b}\frak D_\a W\frak D_{\bar\b}\lbar W-3W\lbar W\),
\ee
where the K\"ahler covariant derivatives are defined as
\be
\frak D_\a\equiv \pa_\a+\pa_\a\ck,\qquad \frak D_{\bar\a}\equiv \pa_{\bar\a}+\pa_{\bar\a}\ck,
\ee
and $L$ is the AdS$_4$ radius.\footnote{Notice that $1/\x L$ corresponds to the gauge coupling, often denoted by $g$ in the supergravity literature.}    
It is also useful to introduce the {\em real} superpotential 
\be\label{real-sup}
\cw=-\frac{\sqrt{2}}{\x L}e^{\ck/2}|W|,
\ee
in terms of which the scalar potential takes the form
\be\label{scalar-potential}
\cv=4\cg^{\a\bar\b}\pa_\a\cw\pa_{\bar\b}\cw-\frac32\cw^2.
\ee

Even after specifying the gauging, i.e. the FI parameters, and the prepotential $F$, there are still two potential ambiguities in specifying the theory completely. From a strict bulk point of view these ambiguities are loosely speaking ``gauge choices'', in the sense that they do not affect physical quantities, but they do change the parameterization of the solutions. Understanding these gauge freedoms, therefore, is important in order to compare different choices in the literature. More importantly, however, these gauge freedoms in the bulk are often lifted by imposing boundary conditions on the fields, and so not all choices are {\it a priory} directly compatible with holography.      

The first gauge freedom is related to a phase factor introduced in the Killing spinor projections, and hence in the BPS equations. This is discussed in \cite{DallAgata:2010ejj}, as well as in \cite{Hristov:2010ri} for the case of purely magnetic solutions (see around eq.~(4.16) there). In \cite{DallAgata:2010ejj}, a constraint for the phase factor is derived and its universal solution in terms of the symplectic sections is obtained, which leads to unambiguous BPS equations, without any additional constraint on the symplectic sections. For purely magnetic solutions the constraint derived in \cite{DallAgata:2010ejj} sets the phase factor to zero, in agreement with the choice made in \cite{Hristov:2010ri}. Below we provide an alternative derivation of these BPS equations through Hamilton-Jacobi (HJ) theory, and so we implicitly treat this phase factor in the same way as \cite{DallAgata:2010ejj}.

The second ambiguity arises in the specification of the symplectic sections $X^\L(z)$ in terms of the physical scalar fields $z^\a$ in the vector multiplets. Since there are $n_V+1$ symplectic sections $X^\L$ but only $n_V$ complex scalars $z^\a$, there is an inherent redundancy in specifying the functions $X^\L(z)$. This redundancy is eliminated by a gauge-fixing condition, that can be visualized as a choice for the embedding of the $n_V$-dimensional complex surface spanned by the physical scalars in the vector multiplets inside the ambient space spanned by the sections $X^\L$. Different embeddings do not affect physical quantities such as the {\em real} superpotential, the scalar potential, the K\"ahler metric and the period matrix (of course up to field redefinitions of the physical scalars), but they do transform non-trivially the K\"ahler potential and the holomorphic superpotential. In appendix \ref{STU-parameterizations} we summarize a number of different embeddings of the STU model scalars that have been used in the literature, and we show explicitly how the $\cn=2$ supergravity quantities defined above transform. This is important for translating known black hole solutions to different parameterizations of the STU model, as well as for understanding the holographic dictionary.

\subsection{Ansatz for static dyonic solutions}
We are interested in static solutions of the $\cn=2$ supergravity action \eqref{action} that can potentially carry both magnetic and electric charge. Such solutions can be parameterized by the Ansatz
\bal\label{Bans}
&ds^2_B=dr^2+e^{2A(r)}\left(-f(r)dt^2+d\s_k^2\right),\qquad k=0,\pm1, \NO\\
&A_B^\L=a^\L(r)\tx dt+p^\L \Big(\int d\th\;\o_k(\th)\Big)\tx d\vf, \qquad z^\a_B=z^\a_B(r), \qquad
\bar z^{\bar \b}_B=\bar z_B^{\bar \b}(r),
\eal
so that the field strengths of the Abelian gauge fields take the form
\be\label{FB}
F_B^\L=\tx dA_B^\L=\dot{a}^\L\;\tx dr\wedge \tx dt+p^\L \o_k(\th)\; \tx d\th\wedge\tx d\vf.
\ee
In this Ansatz $d\s_k^2=d\th^2+\o_k^2(\th)d\vf^2$ is the metric on $\S_k=\{S^2, T^2, H^2\}$ respectively for $k=1,0,-1$, namely
\be
\o_k(\th)=\frac{1}{\sqrt{k}}\sin(\sqrt{k}\;\th)=\left\{\begin{tabular}{cl}
	$\sin\th$, & $k=1$, \\
	$\th$, & $k=0$, \\
	$\sinh\th$, & $k=-1$.
	\end{tabular}\right.
\ee
In the case of $H^2$ the non-compact hyperbolic space must be quotiened by a discrete subgroup of the isometry group, {\it  i.e.}, a Fuchsian group, in order to get a compact Riemann surface of genus $\frak g>1$.

Inserting the Ansatz \eqref{Bans} in the field equations following from the $\cn=2$ supergravity action \eqref{action} results in the following set of coupled equations  
\begin{subequations}\label{BEOM}
	\begin{align}
	& 2\dot A\Big(3\dot A+\frac{\dot{f}}{f}\Big)-\cg_{\a\bar\b}\dot z^\a\dot{\bar z}^{\bar \b}+\cv-2ke^{-2A}+e^{-4A}\ci_{\L\S}\(4e^{2A}f^{-1}\dot a^\L\dot a^\S+p^\L p^\S\)=0,\\
	&\ddot{A}+\dot{A}\Big(3\dot{A}+\frac{\dot{f}}{2f}\Big)+\frac{1}{2}\Big(\cv-2ke^{-2A}+e^{-4A}\ci_{\L\S}\(4e^{2A}f^{-1}\dot{a}^\L\dot a^\S+p^\L p^\S\)\Big)=0, \\
	&\ddot{f}+\dot{f}\Big(3\dot{A}-\frac{\dot{f}}{2f}\Big)+2kfe^{-2A}-2fe^{-4A}\ci_{\L\S}\(4e^{2A}f^{-1}\dot{a}^\L\dot a^\S+p^\L p^\S\)=0, \\
	& 2\cg_{\a\bar\b}\ddot{\bar z}^{\bar\b}_B+2\pa_\g\cg_{\a\bar\b}\dot z_B^{\g}\dot{\bar z}^{\bar\b}_B+2\pa_{\bar\g}\cg_{\a\bar\b}\dot{\bar z}_B^{\bar\g}\dot{\bar z}^{\bar\b}_B-2\pa_\a\cg_{\g\bar\b}\dot z_B^{\g}\dot{\bar z}^{\bar\b}_B+2\cg_{\a\bar\b}\Big(3\dot{A}+\frac{\dot{f}}{2f}\Big)\dot{\bar z}^{\bar\b}_B-\pa_\a \cv\NO\\
	&+e^{-4A}\pa_\a\ci_{\L\S}\(4e^{2A}f^{-1}\dot{a}^\L\dot a^\S-p^\L p^\S\)-4 f^{-1/2}e^{-3A}\pa_\a\car_{\L\S}\dot a^\L p^\S=0,\\
	& 2\cg_{\a\bar\b}\ddot{z}^{\a}_B+2\pa_\g\cg_{\a\bar\b}\dot z^{\g}_B\dot{z}^{\a}_B+2\pa_{\bar\g}\cg_{\a\bar\b}\dot{\bar z}^{\bar\g}_B\dot{ z}^{\a}_B-2\pa_{\bar\b}\cg_{\a\bar\g}\dot z_B^{\a}\dot{\bar z}^{\bar\g}_B+2\cg_{\a\bar\b}\Big(3\dot{A}+\frac{\dot{f}}{2f}\Big)\dot{z}^{\a}_B-\pa_{\bar\b} \cv\NO\\
	&+e^{-4A}\pa_{\bar\b}\ci_{\L\S}\(4e^{2A}f^{-1}\dot{a}^\L\dot a^\S-p^\L p^\S\)-4 f^{-1/2}e^{-3A}\pa_{\bar\b}\car_{\L\S}\dot a^\L p^\S=0,\\
	&\pa_r\Big(2\ci_{\L\S} e^{A}f^{-1/2}\dot{a}^\S-\car_{\L\S}p^\S\Big)=0,
	\end{align}
\end{subequations}
where a dot $\cdot$ denotes a derivative with respect to the radial coordinate $r$.
The last equation, which comes from the Maxwell equation, can be integrated to obtain 
\be\label{charge}
2\ci_{\L\S} e^{A} f^{-1/2}\dot a^\S-\car_{\L\S}p^\S=-q_\L,
\ee
where the integration constants $q_\L$ are electric charges associated with the Abelian gauge fields $A^\L_B$.

\subsection{Effective superpotential and first order equations}
\label{eff-sup}

First order flow equations for static solutions of $\cn=2$ gauged supergravity are known not only for BPS black holes \cite{DallAgata:2010ejj}, but also for several examples of non-extremal black holes  \cite{Ceresole:2007wx,Andrianopoli:2007gt,Andrianopoli:2009je,Trigiante:2012eb,Gnecchi:2012kb,Gnecchi:2014cqa,Klemm:2016wng}. In all these cases, the procedure for deriving the first order equations involves writing the on-shell action as a sum of squares. Although this procedure is sufficiently general for static and spatially homogeneous solutions, in practice only a limited number of flow equations can be obtained this way. HJ theory, however, provides a systematic and general procedure for deriving first order equations, even for non-static and spatially dependent solutions.\footnote{See appendix \ref{ham} for the Hamiltonian formulation of the theory described by the $\cn=2$ Lagrangian \eqref{action}.} For {\em any} solution of the action \eqref{action} these first order equations are given in \eqref{flow}, where the Hamilton principal functional $\cs$ plays the role of a generalized effective superpotential. In particular, HJ theory provides an equation -- the HJ equation -- for the effective superpotential, which can therefore be determined systematically by seeking a solution to the HJ equation. 

For static solutions of the form \eqref{Bans} the general first order equations following from HJ theory were obtained in \cite{Lindgren:2015lia,Klemm:2017pxv}.\footnote{Some of the earlier works, e.g. \cite{Andrianopoli:2009je,Trigiante:2012eb} also employ the HJ method, but only for special cases where the effective superpotential is a function of the scalar fields only. Another approach to first order equations for static black holes was presented in \cite{Kiritsis:2012ma}, but that formulation uses a scalar field as the radial coordinate and amounts to a rewriting of the second order equations of motion. In particular, the first order equations derived in \cite{Kiritsis:2012ma} are strictly on-shell, in the sense that every single solution of the equations of motion is governed by a different effective superpotential. } The result can be summarized as follows: given a solution $\cu(z,\bar z,A)$ of the effective superpotential equation
\be\label{sup-eq}\boxed{
4\cg^{\a\bar\b}\pa_\a\cu\pa_{\bar\b}\cu-\frac12(3+\pa_A)\cu^2=\cv_{eff},}
\ee
where
\be\boxed{
\cv_{eff}=\cv-2ke^{-2A}+e^{-4A}\ci_{\L\S}p^\L p^\S+e^{-4A}\ci^{\L\S}\(q_\L-\car_{\L M}p^M\)\(q_\S-\car_{\S N}p^N\),}
\ee
any solution of the first order equations 
\bal\label{sup-flow-eqs}
\boxed{
\begin{aligned}
&\dot A=-\frac{1}{2}\cu,\qquad 
\frac{\dot f}{f}= -\pa_A\cu,\qquad
\dot z^\a_B=2\cg^{\a\bar\b}\pa_{\bar\b}\cu,\qquad
\dot z^{\bar\b}_B=2\cg^{\a\bar\b}\pa_{\a}\cu,\\
&\dot a^\L=\frac12e^{-A}f^{1/2}\ci^{\L\S}\(\car_{\S M}p^M-q_\S\), 
\end{aligned}
}
\eal
automatically solves the second order equations \eqref{BEOM}.\footnote{One may ask the converse question, namely whether for any solution of the second order equations \eqref{BEOM} there is a superpotential $\cu(z,\bar z, A)$ such that the first order equations \eqref{sup-flow-eqs} hold. This is an interesting and subtle question. Locally in field space this should indeed be the case. Globally, however, a different superpotential $\cu(z,\bar z, A)$ may be necessary in different patches in field space in order to describe a full solution of the second order equations of motion. This happens when e.g. the variables $z$, $\bar z$ and $A$ are not monotonic functions of the radial coordinate. However, for supersymmetric black holes the function $\cu(z,\bar z, A)$ is related to the true superpotential of the theory and so it exist globally in field space.} As was shown in \cite{Lindgren:2015lia} (which focused on the case $k=0$), these flow equations follow from the HJ equation associated with the Hamiltonian constraint in \eqref{constraints}, using the separable ansatz 
\be\label{HJ-sol}\boxed{
	\cs=-\frac{1}{\k^2}\int d^3x\sqrt{\s_k}\Big(e^{3A}f^{1/2}\cu(z,\bar z, A)+2q_\L a^\L\Big).}
\ee
The HJ equation then reduces to the superpotential equation \eqref{sup-eq} for 
the function $\cu(z,\bar z, A)$, and the flow equations \eqref{flow} reduce to the first order equations \eqref{sup-flow-eqs}.

\subsection{Regularized on-shell action}

Given Hamilton's principal function \eqref{HJ-sol} we can easily evaluate the on-shell action with a radial UV cutoff for any solution of the form \eqref{Bans}. We first observe that the only term containing second order derivatives in the Lagrangian \eqref{action} is the bulk Ricci scalar. Using the decomposition \eqref{Ricci} of the bulk Ricci scalar allows one to isolate the terms that contain two derivatives in the radial coordinate. Assuming there is a horizon at $r=r_h$ the on-shell action \eqref{action} evaluated with a radial cutoff $r_o$ takes the form
\be
S_{\rm reg}=\frac{1}{\k^2}\int_{r_h} d^3x\sqrt{-\g}\; K+\int_{r_h}^{r_o} dr \; L,
\ee
where $L$ is the radial Lagrangian \eqref{radial-lagrangian} and the total derivative term from the Ricci scalar evaluated on the cutoff has canceled against the Gibbons-Hawking term. Since the Hamiltonian \eqref{radial-hamiltonian} vanishes on-shell, the regularized on-shell action becomes 
\bal
S_{\rm reg}=&\;\frac{1}{\k^2}\int_{r_h} d^3x\sqrt{-\g}\; K+\int_{r_h}^{r_o} dr \int d^3x\Big(\pi^{ij}\dot{\g}_{ij}+\pi_{\a}\dot{z}^\a+\pi_{\bar\b}\dot{\bar z}^{\bar\b}+\pi_\L^i\dot{A}^\L_i\Big)\NO\\
&\;=\frac{1}{\k^2}\int_{r_h} d^3x\sqrt{-\g}\; K+\int_{r_h}^{r_o} dr \int d^3x\Big(\frac{\d \cs}{\d\g_{ij}}\dot{\g}_{ij}+\frac{\d \cs}{\d z^\a}\dot{z}^\a+\frac{\d \cs}{\d\bar z^{\bar\b}}\dot{\bar z}^{\bar\b}+\frac{\d \cs}{\d A_i^\L}\dot{A}^\L_i\Big)\NO\\
&\;=\frac{1}{\k^2}\int_{r_h} d^3x\sqrt{-\g}\; K+\left.\cs\right|_{r_o}-\left.\cs\right|_{r_h},
\eal
where we have used the expressions \eqref{HJ-momenta} for the canonical momenta. We should point out that this expression for the regularized on-shell action holds for any diffeomorphism invariant two-derivative theory of gravity coupled to matter fields and for any solution of the equations of motion. It follows solely from HJ theory.

For static solutions of the form \eqref{Bans} we have seen that the HJ functional $\cs$ is given by \eqref{HJ-sol}, and so the only term remaining to evaluate is the trace of the extrinsic curvature
\be
K=3\dot A+\frac{\dot f}{2f},
\ee
on the horizon. $\dot A$ vanishes on the horizon, while the blackening factor $f$ behaves as 
\be\label{f-exp}
f=4\p T(u_h-u)+\co(u_h-u)^2,
\ee
where the domain wall coordinate $u$ is related to the radial coordinate $r$ through the definition \cite{Lindgren:2015lia}  
\be
\pa_r=-\sqrt{f}e^{-A}\pa_u.
\ee
It follows that 
\be
\left.e^{3A}f^{1/2}\frac{\dot f}{2f}\right|_{r_h}=-\left.\frac12e^{2A}\pa_u f\right|_{u_h}=2\p T\left.e^{2A}\right|_{u_h}.
\ee
Hence, the {\em Lorentzian} regularized on-shell action for any solution of the form \eqref{Bans} is given by 
\be\label{Sreg}\boxed{
	S_{\rm reg}=-\frac{1}{\k^2}\int_{r_o} d^3x\sqrt{\s_k}\; e^{3A}f^{1/2}\cu+\(a^\L(r_o)-a^\L(r_h)\)Q_\L\int dt+\frac{2\p T}{\k^2}\int dt A_h,}
\ee
where
\be\label{horizon-area}
A_h={\rm Vol}(\S_k)\left.e^{2A}\right|_{h},
\ee
is the area of the horizon,
\be\label{e-charges}
Q_\L\equiv-\frac{2q_\L}{\k^2}{\rm Vol}(\S_k),
\ee
are the electric charges, and ${\rm Vol}(\S_k)$ is the area of the compact surface $\S_k$.

\subsection{Supersymmetric superpotential and BPS equations for dyonic black holes}

The superpotential equation \eqref{sup-eq} admits the exact solution 
\be\label{U-BPS}\boxed{
	\cu\sbtx{BPS}=-\frac{\sqrt{2}}{\x L}e^{\ck/2}|W+ie^{-2A}Z|,}
\ee
where $W$ is the holomorphic superpotential given in \eqref{hol-sup} and $Z$ is the central charge 
\be\label{central-charge}
Z=-\sqrt{2}\;\x L\(p^\L F_\L+q_\L X^\L\),
\ee
provided the magnetic charges satisfy the Dirac quantization condition
\be\label{twist}
p^\L=-\frac{L}{\sqrt{2}}\frak n^\L,\qquad \sum_\L \frak n^\L=2k.
\ee
This is precisely the superpotential obtained in \cite{DallAgata:2010ejj} for dyonic BPS black holes of the $U(1)^4$ gauged supergravity by using the Bogomol'nyi argument of writing the on-shell action as a sum of squares. Our derivation, however, is entirely different, and relies solely on HJ theory. A similar derivation of this superpotential using HJ theory was given in \cite{Klemm:2016wng,Klemm:2017pxv}. The identification of the exact superpotential \eqref{U-BPS} with the true superpotential coming from the supersymmetry variation of the fermionic fields, together with the flow equations \eqref{sup-flow-eqs}, imply that supersymmetric solutions of the action \eqref{action} satisfy the BPS equations
\bal
\label{BPS-eqs}
\boxed{
\begin{aligned}
	\dot A=&\;\frac{1}{\sqrt{2} \x L}e^{\ck/2}|W+ie^{-2A}Z|\;,\\
	\frac{\dot f}{f}=&\;\frac{\sqrt{2}}{\x L}e^{\ck/2}\pa_A|W+ie^{-2A}Z|\;,\\
	\dot z^\a_B=&\;-\frac{\sqrt{2}}{\x L}\ck^{\a\bar\b}\pa_{\bar\b}\(e^{\ck/2}|W+ie^{-2A}Z|\),\\
	\dot z^{\bar\b}_B=&\;-\frac{\sqrt{2}}{\x L}\ck^{\a\bar\b}\pa_{\a}\(e^{\ck/2}|W+ie^{-2A}Z|\),\\
	\dot a^\L=&\;\frac12e^{-A}f^{1/2}\ci^{\L\S}\(\car_{\S M}p^M-q_\S\). 
\end{aligned}}
\eal
Recall that the dots $\cdot$ in these equations denote a derivative with respect to the radial coordinate $r$ defined through the ansatz \eqref{Bans}. 

\paragraph{Near extremal superpotential} The ambiguities in taking extremal limits of black holes are well known  \cite{Gibbons:1994ff,Carroll:2009maa}. It is also understood that to capture certain aspects of extremal black holes, such as the thermodynamics, it is necessary to start from the corresponding non-extremal solutions and approach the extremal ones in a limiting process, as has been done, for example, in computations of the entropy function \cite{Sen:2008yk}. In particular, in order to evaluate the on-shell action for BPS solutions using the regularized expression \eqref{Sreg} it is necessary to evaluate it first on near extremal solutions and then take the extremal limit. The reason for this is that the temperature $T\to 0$ in the extremal limit, while the integral over the Euclidean time gives a factor of $\beta=1/T\to\infty$. Starting with near extremal solutions renders $\b$ finite and $T$ non-zero, leading to an  expression that admits a well defined limit as $T\to0$.

One of the advantages of the HJ method is that it provides an equation for the effective superpotential $\cu$, for both supersymmetric and non-supersymmetric black holes. In order to determine the superpotential for near-extremal black holes, therefore, one can solve \eqref{sup-eq} in perturbation theory around the BPS superpotential. Inserting the near extremal superpotential
\be\label{NE-sup}
\cu=\cu_{\rm BPS}+\ve \D\cu,
\ee
where $\ve$ is the near extremality parameter, in the superpotential equation \eqref{sup-eq} one finds that the first order correction away from extremality satisfies the linear equation 
\be\label{NE-sup-eq}
	4\cg^{\a\bar\b}\pa_\a\cu_{\rm BPS}\pa_{\bar\b}\D\cu+4\cg^{\a\bar\b}\pa_\a\D\cu\pa_{\bar\b}\cu_{\rm BPS}-3\cu_{\rm BPS}\D\cu-\pa_A(\cu_{\rm BPS}\D\cu)=0.
\ee

However, for the purpose of regularizing the extremal limit of the on-shell action we need not solve this equation to determine the functional form of the first order correction $\D\cu$. Instead, it suffices to prove the following two properties: 
\be\label{near-extremal-conditions}
(i)\quad \ve\propto T^\n,\quad \n>1,\qquad (ii)\quad \D\cu=\co(e^{-3r/L})\quad \text{as}\quad r\to\infty.  
\ee
The second property is straightforward to prove. Using the first order equations \eqref{sup-flow-eqs} to replace the superpotential $\cu_{\rm BPS}$ and its derivatives in \eqref{NE-sup-eq} with the radial derivatives of the fields, as well as the asymptotic identities $A\sim r/L$ and $\f\sim 1$ as $r\to\infty$ for asymptotically AdS$_4$ solutions, the linear equation \eqref{sup-flow-eqs} becomes 
\be
\(\pa_r+3/L\)\D\cu=0,
\ee 
which implies condition (ii) in \eqref{near-extremal-conditions}. Condition (i) can be translated to a statement about the near extremal mass. Namely, our analysis in section \ref{holren} implies that  
\be
M-M_{\rm BPS}=\co(T^\n),
\ee
where $\n$ is the same exponent as in condition (i). However, it is known that $\n=2$ for near extremal black holes with an AdS$_2$ near horizon geometry \cite{Almheiri:2016fws}.

\section{Holographic renormalization and the quantum effective action}
\label{holren}

We now have the necessary ingredients in order to construct the holographic dictionary for the theory defined by the bulk action \eqref{action} and appropriate supersymmetric boundary conditions. We will later identify this theory with a sector of the topologically twisted ABJM theory at Chern-Simons level one. We begin this section by determining the boundary counterterms that render the Dirichlet variational problem well posed. We then derive the additional {\em finite} terms required to impose the desired supersymmetric boundary conditions on the scalars. Having determined all necessary boundary terms, we identify the renormalized operators dual to the bulk fields and obtain general expressions for the renormalized partition function and effective action  for any solutions of the form \eqref{Bans}.   

\subsection{Supersymmetric boundary counterterms and boundary conditions}

The solution \eqref{U-BPS} to the superpotential equation implies that the local boundary counterterms compatible with supersymmetry are given by \cite{Papadimitriou:2011qb,An:2017ihs}
\be\label{susy-ct}\boxed{
	S\sbtx{ct}=\frac{1}{\k^2}\int d^3x\sqrt{-\g}\;\cw\(1-\frac k2\Im\(W^{-1}Z_{m}\)R[\g]\),}
\ee
where the real superpotential $\cw$ is defined in \eqref{real-sup} and $Z\sbtx{m}$ denotes the magnetic part of the central charge \eqref{central-charge}, i.e. 
\be\label{m-central-charge}
Z_{m}\equiv-\sqrt{2}\;\x L p^\L F_\L.
\ee
Several comments are in order here. Firstly, to obtain this expression from \eqref{U-BPS} we have Taylor expanded for small $e^{-2A}$ and truncated the resulting expansion to $\co(e^{-2A})$, since higher order terms vanish as the UV cutoff is removed. Moreover, we have covariantized the warp factor by replacing $k e^{-2A}\to R[\g]/2$ and set the electric charges to zero since they contribute terms finite and non-local in the gauge potentials $A_i^\L$. In contrast, the magnetic charges contribute to the divergent terms, but they are local in the gauge potentials and therefore are acceptable as local covariant counterterms. Despite setting the electric charges to zero in \eqref{susy-ct}, we should stress that these local counterterms are valid for any solution of the theory, charged or uncharged, supersymmetric or not, since these counterterms coincide with the asymptotic solution of the HJ equation for any value of the electric and magnetic charges subject to the quantization condition \eqref{twist}. 

A second remark concerns the fact that in the counterterms \eqref{susy-ct} we have included, besides the divergent terms, all finite {\em local} terms dictated by the supersymmetric superpotential \eqref{U-BPS}. This choice of  finite local counterterms renders the boundary term \eqref{susy-ct} invariant under reparameterizations of the symplectic sections $X^\L$ and hence applicable to any parameterization of the STU model. More importantly, as we argue below, this choice is also dictated by supersymmetry.  

In order to write down the possible local finite counterterms it is necessary to pick a specific parameterization of the symplectic sections $X^\L(z)$. From now on we will mostly work in the Pufu-Freedman (PF) parameterization summarized in appendix \ref{STU-parameterizations}, since this parameterization is compatible with supersymmetric boundary conditions and the holographic dictionary, but it is also particularly convenient for discussing the dual theory. Using the Fefferman-Graham expansions of the vector multiplet scalars in the PF parameterization given in \eqref{FG-FP} and decomposing the scalars in real and imaginary part as
\be
z^\a=\cx^\a+i\cy^\a,\qquad \bar z^{\bar\a}= \cx^\a-i \cy^\a,\qquad \cx^\a,\; \cy^\a\in \bb R,
\ee
one can easily conclude that the finite terms in \eqref{susy-ct} are schematically of the form
\be\label{scheme}
(a)\,\,\, \cx^\a \cx^\b \cx^\g,\quad \cx^\a \cx^\b \cy^\g,\quad \cx^\a \cy^\b \cy^\g,\quad \cy^\a \cy^b \cy^\g,\qquad (b)\,\,\, \cx^\a R[\g],\quad \cy^\a R[\g].
\ee 
Were we to impose Dirichlet boundary conditions on all scalars $\cx^\a$ and $\cy^\a$, such terms would correspond to a choice of renormalization scheme, since the induced fields $\cx^\a$ and $\cy^\a$ would be identified with the covariant sources of the dual operators. However, supersymmetry requires that $\cx^\a$ and $\cy^\a$ be quantized in opposite quantizations \cite{Breitenlohner:1982bm} and comparing with the ABJM theory further specifies that the real part $\cx^\a$ of the vector multiplet scalar is dual to dimension one scalar operators, while the imaginary part $\cy^\a$ is dual to dimension two operators \cite{Freedman:2013ryh}. It follows that $\cx^\a$ should satisfy Neumann or mixed boundary conditions, while $\cy^\a$ must satisfy Dirichlet boundary conditions. Hence, the finite terms involving $\cx^\a$ do not correspond to scheme dependence, since $\cx^\a$ is identified with the dual operator instead of its source. As we will see shortly, the freedom of choosing the coefficient of finite local terms of the form \eqref{scheme} involving $\cx^\a$ corresponds instead to the freedom of interpreting the boundary conditions $\cx^\a$ as Neumann or mixed.  

Since the real and imaginary parts of the vector multiplet scalars should satisfy different boundary conditions, it is necessary to formulate the variational problem in terms of $\cx^\a$ and $\cy^\a$. To this end we decompose the scalar  canonical momenta defined in appendix \ref{ham} as  
\be\label{mom-var}
\p_\a\d z^\a+\p_{\bar\b}\d\bar z^{\bar\b}=\p_\a(\d \cx^\a+i \d \cy^\a)+\p_{\bar\b}(\d \cx^{\bar\b}-i\d \cy^{\bar\b})=\p_\a^{\cx}\d \cx^\a+\p^\cy_\a\d \cy^\a,
\ee
and so the canonical momenta conjugate to $\cx^\a$ and $\cy^\a$ are respectively 
\be
\p_\a^{\cx}=(\p_\a+\p_{\bar\a}),\qquad \p^\cy_\a=i(\p_\a-\p_{\bar\a}).
\ee
Using the counterterms \eqref{susy-ct} we then define the renormalized canonical momenta
\bal\label{ren-momenta}
\P^{ij}=\p^{ij}+\frac{\d S\sbtx{ct}}{\d\g_{ij}},\qquad \P^i_\L=\p^i_\L+\frac{\d S\sbtx{ct}}{\d A_i^\L},\qquad \P^\cx_\a=\p^\cx_\a+\frac{\d S\sbtx{ct}}{\d \cx^\a},\qquad \P^\cy_{\a}=\p^\cy_{\a}+\frac{\d S\sbtx{ct}}{\d \bar \cy^{\a}},
\eal
which are associated with the variational principle  
\be\label{Dirichlet}
\d(S_{\rm reg}+S_{\rm ct})=\int d^3x\Big(\P^{ij}\d\g_{ij}+\P^i\d A_i^\L+\P_\a^{\cx}\d \cx^\a+\P^\cy_\a\d \cy^\a\Big).
\ee
This variational principle corresponds to Dirichlet boundary conditions on the scalars $\cy^\a$ and so we must not add any other boundary term that changes the variational problem for these fields. However, we need to add a very specific {\em finite} boundary term in order to impose Neumann or mixed boundary conditions on the scalars $\cx^\a$, while at the same time preserving supersymmetry.  

An important point that is often confusing is that Neumann and mixed boundary conditions can in fact refer to the {\em same} boundary conditions -- one must first specify the Dirichlet theory with respect to which the Neumann boundary conditions are obtained via a Legendre transformation. Different renormalization schemes in the Dirichlet problem correspond to different definitions of what we refer to as the Neumann theory. This should become clear from the general procedure for imposing Neumann or mixed boundary conditions in the renormalized theory \cite{Papadimitriou:2007sj,Caldarelli:2016nni}, which we now review in the context of supersymmetry.

In order to impose generic mixed boundary conditions on $\cx^\a$ we must start with the {\em renormalized} action corresponding to the Dirichlet problem \eqref{Dirichlet}. As we mentioned earlier, this variational problem picks a specific set of finite local counterterms that would correspond to a choice of renormalization scheme, had we imposed Dirichlet boundary conditions. Any choice of such finite terms is in principle acceptable for the Dirichlet problem, unless there are additional constraints, e.g. supersymmetry. At this point it is useful to summarize the results of \cite{An:2017ihs} in relation to supersymmetric boundary conditions for scalar fields. 
\begin{itemize}

\item Finite terms of type (a) in  \eqref{scheme} are Weyl invariant but generically cannot be supersymmetrized individually. As a result, the coefficient of such terms is fixed to the value dictated by the supersymmetric superpotential and  does not correspond to a choice of supersymmetric scheme. This result was shown for a general field theory background in \cite{Freedman:2013ryh} and \cite{An:2017ihs}, but in flat space it is well known that in order to make the vacuum energy zero (as required by supersymmetry) the supersymmetric superpotential must be used as a counterterm.\footnote{There is a caveat to this rule, however, related to the quadratic term in the Taylor expansion of the superpotential around the fixed point. The supersymmetric superpotential can be used as a counterterm iff the coefficient of the quadratic term is proportional to $\D_-$, and not $\D_+$ \cite{Papadimitriou:2004rz}.}     

\item Finite terms of the form (b) in \eqref{scheme} can be made Weyl invariant by replacing the Ricci scalar with the conformal Laplacian, and they can also be supersymmetrized. Therefore, supersymmetry alone does not fix the coefficient of such terms and they correspond to a choice of supersymmetric scheme in the Dirichlet problem. 

\item It was shown in \cite{An:2017ihs} that starting with a supersymmetric Dirichlet problem, the corresponding Neumann problem is supersymmetric as well. This amounts to the statement that the relevant Legendre transformation can be supersymmetrized.

\end{itemize}

Combining these results for supersymmetric Dirichlet and Neumann boundary conditions with the procedure for imposing mixed boundary conditions in the renormalized theory \cite{Papadimitriou:2007sj,Caldarelli:2016nni}, it is straightforward to see how mixed boundary conditions interplay with supersymmetry. Recall that starting with the Dirichlet problem \eqref{Dirichlet}, imposing mixed boundary conditions on the scalars $\cx^\a$ requires adding a finite boundary term of the form\footnote{\label{mixed-general} More general mixed boundary conditions are possible, where the deformation function $v$ is allowed to depend on other fields present, provided Dirichlet boundary conditions are imposed on these fields. In the present theory we could take $v(\cx,\cy,\g_{ij})$. We will comment on such more general mixed boundary conditions below.} (see Table 2 in \cite{Papadimitriou:2007sj}) 
\be\label{LT}
S_{v}=\int d^3x\;\sqrt{-\g}\;J_\a^v \cx^\a+\int d^3x\sqrt{-\g}\;v(\cx),
\ee
where
\be\label{scalar-source-mixed}
J^v_\a\equiv-\frac{1}{\sqrt{-\g}}\P^\cx_\a-\pa_\a v(\cx),
\ee
is identified with the source of the dual scalar operator and $v(\cx)$ is an arbitrary (polynomial) function. Adding this term to \eqref{Dirichlet} leads to the variational principle  
\be\label{Mixed}
\d (S\sbtx{reg}+S_{\rm ct}+S_v)=\int d^3x\(\Big(\P^{ij}+\frac12\(J_\a^v \cx^\a+v(\cx)\)\g^{ij}\Big)\d\g_{ij}+\P_\L^i\d A_i^\L+\P^\cy_\a\d \cy^\a+\cx^\a\d J^v_\a\).
\ee
From these expressions we can draw the following general conclusions.
\paragraph{Finite terms of type (a):} Supersymmetry aside, a choice of scheme in the Dirichlet problem specified by terms of type (a) in \eqref{scheme} is mapped to a shift in the function $v(\cx)$ for mixed boundary conditions. Adding, for example, the finite term 
\be
\int d^3 x\sqrt{-\g}\;\l_{\a\b\g}\cx^\a \cx^\b \cx^\g,
\ee  
where $\l_{\a\b\g}$ are arbitrary constants specifying a choice of scheme in the Dirichlet problem, leads to a shift in the renormalized canonical momenta according to 
\be
\P^{ij}\to \P^{ij}+\frac12\sqrt{-\g}\;\l_{\a\b\g}\cx^\a \cx^\b \cx^\g \g^{ij},\qquad \P_\a^\cx\to\P_\a^\cx+3\sqrt{-\g}\;\l_{\a\b\g}\cx^\b \cx^\g.
\ee
Keeping the scalar source $J_v$ fixed, we see that this shift in the renormalized canonical momenta is equivalent to shifting the function $v(\cx)$ defining the mixed boundary conditions as 
\be
v(\cx)\to v(\cx)+\l_{\a\b\g}\cx^\a\cx^\b \cx^\g.
\ee
Therefore, one can move this type of terms between the counterterms and the function $v(\cx)$ freely, but the total value of this cubic coupling in the renormalized action is fixed and uniquely determined by the scalar boundary conditions, which require $J^v$ to be kept fixed. The same holds also for the other terms of type (a) in \eqref{scheme}, but those terms correspond to a shift $v(\cx)$ by a more general function $\D v(\cx,\cy)$ (see comment in footnote \ref{mixed-general}), and they also modify the momenta $\P_\a^\cy$. 

There are two simple corollaries of this observation. Firstly, {\em marginal} mixed boundary conditions on the scalars can also be viewed as Neumann boundary conditions. In particular, if the function $v(\cx)$ corresponds to a marginal deformation, then it can be entirely absorbed in a choice of scheme for the original Dirichlet problem. Secondly, in combination with the results of \cite{An:2017ihs} summarized above, the {\em total} value of the marginal scalar terms is fixed by supersymmetry, but these terms can still be moved between the counterterms and the function $v(\cx)$. Namely, starting with a supersymmetric Dirichlet problem, i.e. the value of the couplings $\l_{\a\b\g}$ is fixed by the superpotential of the theory, then the corresponding Neumann boundary condition is supersymmetric, but any mixed boundary condition breaks supersymmetry. However, starting with a generic value of the cubic couplings in the Dirichlet problem such that supersymmetry is broken, then the corresponding Neumann problem is not supersymmetric, but a very specific mixed boundary condition is. Therefore, with our choice to include the full supersymmetric superpotential in the counterterms as in \eqref{susy-ct}, we have to impose Neumann boundary conditions on the scalars $\cx^\a$ since any mixed boundary condition will break supersymmetry. Hence, supersymmetry dictates that starting with the counterterms \eqref{susy-ct}, we must set the function $v(\cx)$ to zero.        

Finally, an interesting situation arises specifically for the finite local terms of the form $\cx^\a\cy^\b\cy^\g$. Since such terms are linear in the scalars $\cx^\a$, they drop out of the Legendre transform with respect to $\cx^\a$. This can be seen immediately from \eqref{LT} and \eqref{scalar-source-mixed} by taking $v(\cx,\cy)=\l_{\a\b\g}\cx^\a\cy^\b\cy^\g$, or equivalently, by changing the scheme by such a term in the original Dirichlet problem. It follows that starting with the value of this coupling in the Dirichlet problem dictated by the superpotential of the theory, such that the Dirichlet problem is supersymmetric, adding a term of the form $\l_{\a\b\g}\cx^\a\cy^\b\cy^\g$ {\em and} performing the Legendre transform with respect to $\cx^\a$ trivially preserves supersymmetry since the final result of the Legendre transformation is identical with that obtained from the original supersymmetric Dirichlet problem. This is precisely the observation made in section 3.8 of \cite{Freedman:2016yue}. However, adding such a term is not trivial because it changes the definition of the scalar source $J_\a$, which affects the calculation of physical observables, e.g. correlation functions. Indeed, it was by looking at the three-point function $\<\co_{\D=1}\co_{\D=2}\co_{\D=2}\>$ that the authors of \cite{Freedman:2016yue} were able to determine the correct value of this cubic coupling.

\paragraph{Finite terms of type (b):} The above observations apply to terms of type (b) in \eqref{scheme} as well, except that such terms correspond to a choice of supersymmetric scheme in the Dirichlet problem, and so supersymmetry does not fix their overall coefficient. Moreover, in contrast to the analogous terms in four dimensions considered in \cite{An:2017ihs}, in three dimensions the type (b) terms involving the scalars $\cx^\a$ are linear in $\cx^\a$ and so they drop out of the Legendre transform, exactly as the terms $\cx^\a\cy^\b\cy^\g$ we just discussed. It follows that such terms trivially preserve supersymmetry in the Legendre transformed theory, but they do affect the definition of the sources $J_\a$ of the dimension one operators and, hence, some argument is required in order to fix the coefficients of such terms. 

Such an argument is provided by the requirement that the sources of the dimension one operators, namely
\be\label{scalar-source-N} 
J_\a\equiv-\frac{1}{\sqrt{-\g}}\P^\cx_\a,
\ee 
vanish on supersymmetric solutions. Notice that we choose to impose Neumann boundary conditions since, as in the case of type (a) terms, mixed boundary conditions can be traded for a choice of scheme in the original Dirichlet problem, without affecting the actual value of the source $J_\a$. Terms of type (b) in the counterterms contribute a constant multiple of the Ricci curvature $R[\g]$ in the renormalized canonical momenta $\P^\cx_\a$. Hence, for solutions with $k=\pm 1$ the sources $J_\a$ are shifted by a constant non-zero term. Requiring that the BPS equations coincide with the condition of vanishing scalar sources, i.e. $\left.J_\a\right|_{\rm BPS}=0$, unambiguously determines the coefficients of the type (b) terms proportional to $\cx^\a$ in the counterterms to be the ones given in \eqref{susy-ct}. This amounts to including the full supersymmetric effective superpotential for magnetic BPS solutions in the counterterms.\footnote{It may be useful to point out that \cite{Freedman:2013ryh} sets the coefficients of the finite terms of type (b) in \eqref{scheme} to zero in the choice of supersymmetric counterterms, while \cite{Halmagyi:2017hmw} does not specify them arguing that they drop out of the Legendre transform. Although the coefficient of the finite terms $\cy^\a R[\g]$ is indeed a choice of supersymmetric scheme, since Dirichlet boundary conditions are imposed on $\cy^\a$, we have argued that the coefficient of the terms $\cx^\a R[\g]$ is in fact determined by the value of the source of the dimension one operators in BPS solutions. Despite the fact that, as correctly pointed out in \cite{Halmagyi:2017hmw}, supersymmetry is (trivially) preserved for any value of the coefficient of $\cx^\a R[\g]$ because such terms cancel out in the Legendre transform, demanding that the single trace source of the dimension one operators in BPS solutions vanishes uniquely determines the coefficients of $\cx^\a R[\g]$ in the supersymmetric counterterms to be those dictated by the supersymmetric superpotential, as in \eqref{susy-ct}.} Notice that we are able to use this argument to determine the coefficient of the finite terms proportional to the Ricci curvature in the counterterms because we are considering BPS solutions with a non-zero boundary curvature, i.e. $k=\pm 1$. As we pointed out earlier, these terms are analogous to the terms $\cx^\a\cy^\b\cy^\g$ discussed in \cite{Freedman:2016yue}, except that those terms contribute to the sources $J_\a$ a term proportional to $\cy^\b\cy^\g$, i.e. to the square of the sources of the dimension two operators. Since the sources $\cy^\a$ vanish in the background solutions considered in \cite{Freedman:2016yue}, this shift in the source $J_\a$ is not visible in the BPS equations, which is why the authors of \cite{Freedman:2016yue} have to use the scalar three-point functions to determine the coefficient of this term.

\subsection{Renormalized holographic observables}

The outcome of the analysis in the previous subsection is that the renormalized generating function in the dual supersymmetric theory is given by  
\be\label{Sren}\boxed{
	\bb W\big[g_{(0)ij},A_{(0)i}^\L,\cy_{-}^\a,J^{+}_{\a}\big]\equiv S\sbtx{ren}=\lim_{r_o\to\infty}\(S\sbtx{reg}+S\sbtx{ct}+S_{v=0}\),}
\ee
where $S_{\rm reg}$ is the regularized on-shell action, including the Gibbons-Hawking term, $S_{\rm ct}$ are the boundary counterterms defined in \eqref{susy-ct}, and $S_{v=0}$ is the Legendre transform \eqref{LT}, with the function $v(\cx)$ set to zero. This generating function depends on the sources 
\bal\label{sources}
g_{(0)ij}=&\;\lim_{r\to\infty} (e^{-2r/L}\g_{ij}),\qquad A_{(0)i}^\L=\;\lim _{r\to\infty} A_{i}^\L,\NO\\
\cy^\a_{-}=&\;\lim_{r\to\infty} (e^{r/L}\cy^\a),\qquad J_\a^{+}=\;\lim_{r\to\infty} (e^{2r/L}J_\a^{v=0}).
\eal

Differentiating $\bb W\big[g_{(0)ij},A_{(0)i}^\L,\cy_{-}^\a,J^{+}_{\a}\big]$ with respect to these sources gives the corresponding one-point functions in the presence of sources. Namely, 
\bal\label{ops}
\<\ct^{ij}\>=&\;\lim_{r\to\infty}\(e^{2r/L}\Big(\frac{2}{\sqrt{-\g}}\P^{ij}+J_\a^{v=0} \cx^\a\g^{ij}\Big)\)=\frac{2}{\sqrt{-g_{(0)}}}\frac{\d\bb W}{\d g_{(0)ij}},\NO\\
\<\cj^i_\L\>=&\;\lim_{r\to\infty}\(\frac{1}{\sqrt{-\g}}\P^{i}_\L\)=\frac{1}{\sqrt{-g_{(0)}}}\frac{\d\bb W}{\d A^\L_{(0)i}},\NO\\
\<\co_\a^{\D=2}\>=&\;\lim_{r\to\infty}\(e^{-r/L}\frac{1}{\sqrt{-\g}}\p^{\cy}_\a\)=\frac{1}{\sqrt{-g_{(0)}}}\frac{\d\bb W}{\d \cy^\a_{-}},\NO\\ 
\<\co^\a_{\D=1}\>=&\;\lim_{r\to\infty}\(e^{r/L}\cx^\a\)=\cx^\a_{-}=\frac{1}{\sqrt{-g_{(0)}}}\frac{\d\bb W}{\d J_\a^{+}}.
\eal
These one-point functions satisfy the following Ward identities, which can be deduced from the first class constraints \eqref{constraints} in the radial Hamiltonian formulation of the bulk dynamics \cite{Caldarelli:2016nni}
\bsub
\bal
&-D_{(0)}^j\<\ct_{ij}\>+\<\co^\a_{\D=1}\>\pa_i J^{+}_\a+\<\co^\a_{\D=2}\>\pa_i \cy^\a_{-}+\(\<\cj^j_\L\>+\frac{2}{\k^2}\e_{(0)}^{jkl}\car_{\L\S}(0)F^\S_{(0)kl}\) F_{(0)ij}^\L=0,\\
&D_{(0)i}\<\cj^i_\a\>=0,\\\
&-\<\ct^i_i\>+2J^{+}_\a\<\co^\a_{\D=1}\>+\cy_{-}^\a\<\co_\a^{\D=2}\>=0.
\eal
\esub

The Legendre transform of the generating functional \eqref{Sren} with respect to any of the sources is the quantum effective action for the vacuum expectation value (VEV) of the corresponding operator. As we will show in the next section, the entropy functional for BPS black holes of the supergravity action \eqref{action} is related to the effective action obtained by Legendre transforming the generating functional \eqref{Sren} with respect to the source $J_\a^{+}$ of the dimension one scalar operators. The resulting quantum effective action is a functional of the VEVs  $\cx^\a_{-}=\<\co^\a_{\D=1}\>$ and takes the form
\be\label{eff-action}\boxed{
\G[(g_{(0)ij},A_{(0)i}^\L,\cy_{-}^\a,\cx^\a_{-}\big)]=\bb W\big[g_{(0)ij},A_{(0)i}^\L,\cy_{-}^\a,J^{+}_{\a}\big]-\int \hskip-.1cm d^3x\; \cx_{-}^\a J_{\a}^{+}=\lim_{r_o\to\infty}\hskip-.1cm\(S\sbtx{reg}+S\sbtx{ct}\).}
\ee
Notice that the Legendre transform simply removes the term $S_{v=0}$ in the generating function and so the effective action of the Neumann theory coincides with the generating function of the Dirichlet one \cite{Papadimitriou:2007sj}. Earlier computations of this effective action, up to two derivatives in the derivative expansion, appeared for a number of different examples in  \cite{deHaro:2006ymc,Papadimitriou:2007sj,Kiritsis:2014kua,Caldarelli:2016nni}.

\subsection{The BPS limit and black hole thermodynamics}

For solutions of the form \eqref{Bans} all the renormalized observables can be related to the corresponding effective superpotential $\cu$. Using the counterterms \eqref{susy-ct} and the expression \eqref{Sreg} for the regularized on-shell action one finds that the renormalized partition function is given by
\bal\label{Sren-ansatz}
	\bb W=&\;-\frac{1}{\k^2}\int_{r_o\to\infty} \hskip-0.6cmd^3x\sqrt{\s_k}\; e^{3A}f^{1/2}(\cu-\cu^{q=0}_{\rm BPS})+\m^\L Q_\L\int dt+\frac{2\p T}{\k^2}\int dt A_h\NO\\
	&\hskip0.1cm+\frac{1}{\k^2}\int_{r_o\to\infty} \hskip-0.6cmd^3x\sqrt{\s_k}\;e^{3A}f^{1/2}\cx^\a\frac{\pa}{\pa\cx^\a}(\cu-\cu^{q=0}_{\rm BPS}),
\eal
where $\cu^{q=0}_{\rm BPS}$ stands for the supersymmetric superpotential \eqref{U-BPS}, with the electric charges set to zero. Moreover, the area of the horizon $A_h$ and the electric charges $Q_\L$ were defined respectively in \eqref{horizon-area} and \eqref{e-charges}, and we have introduced the electric chemical potentials
\be
\label{chemical-potentials}
\m^\L\equiv a^\L(\infty)-a^\L(r_h).
\ee 
Notice that the last term in \eqref{Sren-ansatz} corresponds to the Legendre transform $S_{v=0}$ and, therefore, the effective action \eqref{eff-action} becomes
\be\label{eff-pot}
\G=-\frac{1}{\k^2}\int_{r_o\to\infty} \hskip-0.5cmd^3x\sqrt{\s_k}\; e^{3A}f^{1/2}(\cu-\cu^{q=0}_{\rm BPS})+\m^\L Q_\L\int dt+\frac{2\p T}{\k^2}\int dt\; A_h.
\ee

The one-point functions \eqref{ops} can also be evaluated in terms of the effective superpotential. The general expressions for the one-point functions for any background of the form \eqref{Bans} are given in eq.~(3.34) of \cite{Lindgren:2015lia}. Using the supersymmetric counterterms \eqref{susy-ct} these expressions become
\bal\label{VEVs-ansatz}
\<\ct_{tt}\>=&\;\frac{1}{\k^2}\lim_{r\to\infty}e^{3A}(\cu-\cu_{\rm BPS}^{q=0}),\NO\\
\<\ct_{aa}\>=&\;-\frac{1}{\k^2}\lim_{r\to\infty}e^{3A}\Big(1+\frac12\pa_A\Big)(\cu-\cu_{\rm BPS}^{q=0}),\NO\\
\<\cj^i_\L\>=&\;-\frac{2}{\k^2}q_\L\d^{it},\NO\\
\<\co_\a^{\D=2}\>=&\;-\frac{1}{\k^2}\lim_{r\to\infty}e^{2r/L}\frac{\pa}{\pa\cy^\a}\(1-\cx^\a\frac{\pa}{\pa\cx^\a}\)(\cu-\cu_{\rm BPS}^{q=0}),\NO\\ 
\<\co^\a_{\D=1}\>=&\;\lim_{r\to\infty}\big(e^{r/L}\cx^\a\big).
\eal

\paragraph{The extremal limit} These quantities can be evaluated explicitly in the extremal limit, corresponding to the exact superpotential \eqref{U-BPS}. As we pointed out earlier, in order to evaluate some of these observables in the extremal limit, it is necessary to start from near extremal solutions and take the zero temperature limit in the end. Since all observables are expressed in terms of a generic effective superpotential $\cu$, evaluating them on near extremal solutions amounts to using the near extremal superpotential \eqref{NE-sup}. For large radial cutoff $r_o$ this can be expanded to obtain
\bal
\cu-\cu^{q=0}_{\rm BPS}=&\;\cu_{\rm BPS}-\cu^{q=0}_{\rm BPS}+\ve\D\cu\NO\\
=&\;-\frac{\sqrt{2}}{\x L}e^{\ck/2}|W|\(-e^{-2A}\Im(W^{-1}Z_e)+\co(e^{-4A})\)+\ve\D\cu,
\eal
where $Z_{e}$ denotes the electric part of the central charge \eqref{central-charge}, i.e. 
\be\label{e-central-charge}
Z_{e}\equiv-\sqrt{2}\;\x L q_\L X^\L.
\ee

Using this expansion, and the Pufu-Freedman parameterization of the STU model discussed in appendix \ref{STU-parameterizations}, we find that the effective action \eqref{eff-pot} for near extremal black holes takes the form
\bal\label{eff-pot-NE}
\G=&\;\frac{2}{\k^2}\int d^3x\sqrt{\s_k}\;(\frak m^\L-\m^\L)q_\L+\frac{2\p T}{\k^2}\int dt\; A_h-\frac{1}{\k^2}\int_{r_o\to\infty} \hskip-0.5cmd^3x\sqrt{\s_k}\; e^{3A}f^{1/2}\ve\D\cu,
\eal
where
\be\label{m-Y}
\frak m^0\equiv\frac{1}{8}(\cy^1_{-}+\cy^2_{-}+\cy^3_{-}),\quad
\frak m^\a\equiv-\frac{1}{8}\((-1)^{\d_{\a 1}}\cy^1_{-}+(-1)^{\d_{\a 2}}\cy^2_{-}+(-1)^{\d_{\a 3}}\cy^3_{-}\),\quad \a=1,2,3.
\ee
Notice that 
\be\label{mass-sum}
\sum_\L\frak m^\L=0.
\ee
Moreover, the second property in \eqref{near-extremal-conditions} of the near extremal superpotential ensures that the last term in \eqref{eff-pot-NE}, which is proportional to $\D\cu$, is finite as the cutoff is removed at fixed $\ve$. Moreover, the first result in \eqref{near-extremal-conditions} implies that as $\ve\to 0$, this term gives a zero contribution to the Euclidean effective action, and so only the first two terms in \eqref{eff-pot-NE} can potentially contribute in the extremal limit. The term involving the area of the horizon has a finite extremal limit, but the integrand of the first term in \eqref{eff-pot-NE} is not proportional to the temperature and so it seems to lead to a divergent contribution to the extremal Euclidean effective action due to the infinite periodicity of the Euclidean time, i.e. $\b\to\infty$. We therefore conclude that supersymmetric solutions must satisfy the boundary condition        
\be\label{susy-sources}\boxed{
\frak m^\L=\m^\L.}
\ee
This condition relates the sources $\cy_{-}^\a$ of the dimension two scalar operators to the electric chemical potentials and, therefore, is an additional requirement for the Dirichlet boundary conditions on the scalars $\cy^\a$ and the gauge fields $A_i^\L$ to be supersymmetric. Provided the condition \eqref{susy-sources} holds, therefore, the effective action for BPS solutions is given by the area of the horizon, namely 
\be\label{extremal-eff-pot}\boxed{
\G_{\rm BPS}[\cx_{-}^\a;\frak m^\L,\frak n^\L]=\frac{2\p T}{\k^2}\int dt\; A_h[\cx_{-}^\a;\frak m^\L,\frak n^\L],}
\ee
where we have kept the temperature as a regulator in the off-shell effective action. It is only after Wick rotation to Euclidean signature that the temperature will cancel against the perimeter of the Euclidean time circle. It should be stressed that at this point the horizon area is not equal to the extremal entropy since it is evaluated at {\em arbitrary} VEVs $\cx^\a_{-}=\<\co^\a_{\D=1}\>$. $\G_{\rm BPS}[\cx_{-}^\a;\frak m^\L,\frak n^\L]$ is the field theory quantum effective action for these VEVs. As we will show in the next section, the extremization of this effective action, at fixed magnetic fluxes $\frak n^\L$, is the field theory realization of the attractor mechanism in the bulk. 

The area of the horizon can be evaluated explicitly for BPS black holes using the exact superpotential \eqref{U-BPS}. This is because the effective superpotential vanishes on the horizon \cite{DallAgata:2010ejj}, i.e.\footnote{The fact that the effective superpotential $\cu$ vanishes on the horizon, even for non-BPS black holes, follows from the first order equations \eqref{sup-flow-eqs} and the near horizon behavior of the blackening factor in \eqref{f-exp}.} 
\be
\left.\cu_{\rm BPS}\right|_{h}=0,
\ee
and therefore 
\be
\left.e^{2A}\right|_{h}=-\left.i W^{-1}Z\right|_h=i\sqrt{2}\;\x L\left.W^{-1}(p^\L F_\L+q_\L X^\L)\right|_h.
\ee
Inserting this expression for the warp factor in \eqref{extremal-eff-pot} we arrive at the following general expression for the effective action of dyonic BPS black holes of the $U(1)^4$ gauge supergravity: 
\be\label{extremal-eff-pot-explicit}\boxed{
	\G_{\rm BPS}[\cx_{-}^\a;\frak m^\L,\frak n^\L]=\frac{2\p T}{\k^2}\int dt\; \int d^2x\sqrt{\s_k}\; \left.i\sqrt{2}\;\x LW^{-1}(p^\L F_\L+q_\L X^\L)\right|_h.}
\ee
In the next section we will evaluate this effective action explicitly, first on magnetic and then dyonic BPS black holes. Notice that this effective action depends on the UV parameters $\cx_{-}^\a$, $\frak m^\L$, $\frak n^\L$ and, therefore, it is necessary to know the full black hole solutions to correctly evaluate it. In particular, it is {\em not} sufficient to evaluate the effective action using the near horizon solutions, since this does not determine the relation between the parameters of the near horizon solutions to the physical UV modes $\cx_{-}^\a$, $\frak m^\L$ and $\frak n^\L$.   

Using the exact superpotential \eqref{U-BPS}, we can also evaluate the one-point function
\eqref{VEVs-ansatz} for BPS black holes. A straightforward calculation determines
\bal
\label{extremal-VEVs}
\boxed{
\begin{aligned}
&\<\ct_{tt}\>=-\frac{2}{\k^2}\frak m^\L q_\L,\qquad \<\ct_{ta}\>=\<\ct_{ab}\>=0,\qquad \<\cj^i_\L\>=-\frac{2}{\k^2}q_\L\d^{it},\\
&\<\co_\a^{\D=2}\>=\frac{1}{4\k^2}\Big(q_0-(-1)^{\d_{\a 1}}q_1-(-1)^{\d_{\a 2}}q_2-(-1)^{\d_{\a 3}}q_3\Big),\qquad \<\co^\a_{\D=1}\>=\cx^\a_{(0)}.
\end{aligned}
}
\eal
These expressions for the one-point functions of BPS solutions have a number of important consequences. Firstly, using the definition of the fermion masses $\frak m^\L$ in \eqref{m-Y} and the VEVs of the dimension two operators in \eqref{extremal-VEVs} we deduce that  
\be
\<\co_\a^{\D=2}\>\d\cy^\a_{-}=\frac{2}{\k^2}q_\L\d\frak m^\L.
\ee
In combination with the effective action \eqref{eff-pot-NE} this result implies that 
\be
\frac{\pa}{\pa\frak m^\L}\int dt\; A_h[\cx_{-}^\a;\frak m^\L,\frak n^\L]=0,
\ee
or equivalently
\be\label{extremum-Y}
\frac{\pa}{\pa\cy^\a_-}\int dt\; A_h[\cx_{-}^\a;\frak m^\L,\frak n^\L]=0,
\ee
and so the BPS effective action is extremized with respect to the sources of the dimension two operators, with the extremal values given by the chemical potentials as in \eqref{susy-sources}. This observation will play a central role in our field theory interpretation of the attractor mechanism. 

Another implication of the supersymmetric one-point functions \eqref{extremal-VEVs} is that the supersymmetric mass of dyonic BPS black holes is \cite{Papadimitriou:2005ii,Caldarelli:2016nni}
\be\boxed{
M=-\frac{2}{\k^2}\frak m^\L q_\L{\rm Vol}(\S_k),}
\ee
and, as a direct consequence of the relation \eqref{susy-sources}, satisfies the BPS relation
\be\label{BPS-relation}\boxed{
M-\m^\L Q_\L=0.}
\ee

Finally, collecting the above results, we can evaluate the Euclidean on-shell action, which is holographically identified with the grand canonical potential, i.e. the Gibbs free energy, for dyonic BPS black holes: 
\be\label{Gibbs}\boxed{
\ci=-\bb W^{\rm E}_{\rm BPS}=-\G^{\rm E}_{\rm BPS}=-S,}
\ee
where $S$ is the extremal entropy, evaluated at the extremum of the effective action.
This result agrees with that obtained in \cite{Halmagyi:2017hmw}, as well as \cite{Azzurli:2017kxo} in the case of AdS$_4$ black holes without scalars. Our derivation, however, provides an explicit proof that, as anticipated in \cite{Halmagyi:2017hmw}, the free energy of extremal asymptotically AdS black holes 
is a direct consequence of imposing the BPS relation \eqref{BPS-relation} in the quantum statistical relation for general asymptotically AdS black holes \cite{Papadimitriou:2005ii}
\be
\ci=\b(M-ST-\m_\L Q^\L),
\ee 
and taking the extremal limit.

\section{Holographic attractor mechanism and the entropy functional}
\label{attractor}

In this section we will demonstrate that extremizing the holographic quantum effective action \eqref{extremal-eff-pot-explicit} for BPS black holes  with respect to the VEVs of the dimension one scalar operators determines the correct supersymmetric values for these VEVs. Moreover, we will show that the value of the effective action at the extremum coincides with the black hole entropy. This provides a purely field theoretic extremization principle, which we dub {\em the holographic attractor mechanism}. 

Evaluating the effective action \eqref{extremal-eff-pot-explicit} explicitly as a function of the UV parameters $\cx_{-}^\a$, $\frak m^\L$ and $\frak n^\L$ is a formidable task: it requires knowledge of the general -- not necessarily regular in the interior -- solution of the BPS equations, with arbitrary $\cx_{-}^\a$, in closed form. This is necessary in order to express explicitly the area of the horizon in terms of the arbitrary scalar VEVs $\cx_{-}^\a$ at the UV. However, we will see that knowledge of this solution is not necessary in order to obtain the extremal entropy. In particular, we will demonstrate by means of a concrete example that extremizing the effective action with respect to the scalar VEVs is equivalent to extremizing the expression \eqref{extremal-eff-pot-explicit} for the effective action with respect to the values of the physical scalars on the horizon. If one knowns the exact BPS black hole solution corresponding to the extremum of the effective potential, then the values of the UV VEVs at the extremum can be determined as well.       

\paragraph{The BPS swampland} The metric ansatz \eqref{Bans} is designed so that the first order equations \eqref{sup-flow-eqs} take the simplest form. However, to obtain explicit black hole solutions it is convenient to reparameterize the metric by defining  
\be\label{Bans-map}
e^{2A}=\bar r^2h(\bar r),\qquad
f=\frac{b(\bar r)}{\bar r^2 h^2(\bar r)},\qquad
dr=\frac{h^{\frac12}(\bar r)}{b^{\frac12}(\bar r)}d\bar r.
\ee  
The ansatz \eqref{Bans} then becomes
\bal\label{Bans-update}
ds^2=&\;h(\bar r)b^{-1}(\bar r)d\bar r^2-h^{-1}(\bar r)b(\bar r)dt^2+h(\bar r)\bar r^2d\s_k^2,\NO\\
A^\L=&\;a^\L(\bar r)\tx dt+p^\L \Big(\int d\th\;\o_k(\th)\Big)\tx d\vf,\qquad z^\a=z^\a(\bar r),
\eal
while the BPS equations \eqref{BPS-eqs} take the form
\bal
\label{BPS-eqs-update}
		b^{\frac12}(\bar r)h^{-\frac12}(\bar r)A'=&\;\frac{1}{\sqrt{2} \x L}e^{\ck/2}|W+ie^{-2A}Z|\;,\NO\\
		b^{\frac12}(\bar r)h^{-\frac12}(\bar r)\frac{f'}{f}=&\;\frac{\sqrt{2}}{\x L}e^{\ck/2}\pa_A|W+ie^{-2A}Z|\;,\NO\\
		b^{\frac12}(\bar r)h^{-\frac12}(\bar r)z'^\a=&\;-\frac{\sqrt{2}}{\x L}\ck^{\a\bar\b}\pa_{\bar\b}\(e^{\ck/2}|W+ie^{-2A}Z|\),\NO\\
		b^{\frac12}(\bar r)h^{-\frac12}(\bar r)\bar z'^{\bar\b}=&\;-\frac{\sqrt{2}}{\x L}\ck^{\a\bar\b}\pa_{\a}\(e^{\ck/2}|W+ie^{-2A}Z|\),\NO\\
		 2\bar r^2 h(\bar r)a'^\L=&\;\ci^{\L\S}\(\car_{\S M}p^M-q_\S\),
\eal
where ${}'$ denotes a derivative with respect to the radial coordinate $\bar r$.

Since the BPS equations \eqref{BPS-eqs-update} are first order, their general solution contains $n_V+3$ real and $n_V$ complex integration constants, i.e. one for each equation. One of the real integration constants is related to rescaling of the radial coordinate $\bar r$ and is fixed by requiring that the solution is asymptotically AdS$_4$ with radius $L$. A second real integration constant is fixed by a suitable regularity condition in the interior, such as the existence of a smooth horizon. Moreover, as we have seen in the previous section in eq.~\eqref{susy-sources}, supersymmetry relates the $n_V+1$ electric chemical potentials to the $n_V$ sources of the dimension two operators. The general supersymmetric asymptotically AdS$_4$ solution is therefore parameterized by $2n_V$ real integration constants: $n_V$ independent electric chemical potentials and $n_V$ VEVs for the dimension one operators. The chemical potentials, however, are a boundary condition and can therefore be set to any desired value. The VEVs of the dimension one operators, on the other hand, are dynamically determined by the theory. Namely, they are fixed by extremizing the quantum effective action, evaluated on the {\em BPS swampland}, i.e. the general supersymmetric solution with $n_V$ arbitrary scalar VEVs, for given chemical potentials.

\subsection{Magnetic BPS black holes}

For real scalars and vanishing electric charges the BPS equations \eqref{BPS-eqs-update} are sufficiently simple to be written explicitly. In the Pufu-Freedman parameterization, discussed in appendix \ref{STU-parameterizations}, they take the form
\bal
\label{BPS-eqs-update-magnetic}
	b^{\frac12}(\bar r)h^{-\frac12}(\bar r)A'=&\;\frac{1}{4L\sqrt{(1-(z^1)^2)(1-(z^2)^2)(1-(z^3)^2)}}\Big[4\(1+z^1z^2z^3\)\NO\\
	&\hskip-1.5cm-\sqrt{2}\; L e^{-2A}\Big(p^0(1-z^1)(1-z^2)(1-z^3)+p^1(1-z^1)(1+z^2)(1+z^3)\NO\\
	&\hskip-1.5cm+p^2(1+z^1)(1-z^2)(1+z^3)+p^3(1+z^1)(1+z^2)(1-z^3)\Big)\Big],\NO\\
	b^{\frac12}(\bar r)h^{-\frac12}(\bar r)\frac{f'}{f}=&\;\frac{\sqrt{2}\; e^{-2A}}{\sqrt{(1-(z^1)^2)(1-(z^2)^2)(1-(z^3)^2)}}\Big(p^0(1-z^1)(1-z^2)(1-z^3)\NO\\
	&\hskip-1.5cm+p^1(1-z^1)(1+z^2)(1+z^3)+p^2(1+z^1)(1-z^2)(1+z^3)+p^3(1+z^1)(1+z^2)(1-z^3)\Big),\NO\\
	b^{\frac12}(\bar r)h^{-\frac12}(\bar r)z'^\a=&\;-\frac{(1-(z^\a)^2)}{4L\sqrt{(1-(z^1)^2)(1-(z^2)^2)(1-(z^3)^2)}}\Big[4\(z^\a+z^1z^2z^3/z^\a\)\NO\\
	&\hskip-1.5cm+\sqrt{2}\; L e^{-2A}\Big(p^0(1-z^1)(1-z^2)(1-z^3)-(-1)^{\d_{\a 1}}p^1(1-z^1)(1+z^2)(1+z^3)\NO\\
	&\hskip-1.5cm-(-1)^{\d_{\a 2}}p^2(1+z^1)(1-z^2)(1+z^3)-(-1)^{\d_{\a 3}}p^3(1+z^1)(1+z^2)(1-z^3)\Big)\Big],\NO\\
	a'^\L=&\;0. 
\eal

\paragraph{General magnetic BPS solution} The general solution of the BPS equations \eqref{BPS-eqs-update-magnetic} can be sought in the form of an asymptotic expansion in the UV. In particular, ensuring that the solution is asymptotically AdS$_4$ with radius $L$, we write 
\bal\label{magnetic-uv-exp}
h(\bar r)=&\;\sqrt{1+h^{(1)} L/\bar r+h^{(2)}(L/\bar r)^2+\cdots}\;,\NO\\
b(\bar r)=&\;\frac{\bar r^2}{L^2}+b^{(-1)}\bar r/L+b^{(0)}+b^{(1)}L/\bar r+b^{(2)}(L/\bar r)^2+\cdots,\NO\\
v_\a(\bar r)=&\;1+v_\a^{(1)}L/\bar r+v_\a^{(2)}(L/\bar r)^2+v_\a^{(3)}(L/\bar r)^3+\cdots,
\eal
where the functions $v_\a$ determine the scalars $z^\a$ through the relations 
\be\label{z-v}
z^1=\frac{1-v_1}{1+v_1},\qquad z^2=\frac{1-v_2}{1+v_2},\qquad z^3=\frac{1-v_3}{1+v_3}.
\ee

Inserting these expansions in the BPS equations \eqref{BPS-eqs-update-magnetic} we find up to the first two subleading orders
\bal
&b^{(-1)}=h^{(1)},\quad b^{(0)}=h^{(2)}+k,\quad h^{(2)}=\frac18\Big(3(h^{(1)})^2-4\sum_\a (v_\a^{(1)})^2\Big),\NO\\
&v_1^{(2)}=\frac14\Big(4k-4\frak n_2-4\frak n_3-h^{(1)} v_1^{(1)}+2(v_1^{(1)})^2-2v_2^{(1)}v_3^{(1)}\Big),\NO\\
&v_2^{(2)}=\frac14\Big(4k-4\frak n_1-4\frak n_3-h^{(1)} v_2^{(1)}+2(v_2^{(1)})^2-2v_1^{(1)}v_3^{(1)}\Big),\NO\\
&v_3^{(2)}=\frac14\Big(4k-4\frak n_2-4\frak n_1-h^{(1)} v_3^{(1)}+2(v_3^{(1)})^2-2v_2^{(1)}v_1^{(1)}\Big).
\eal
We have determined these expansions up to the terms $h^{(5)}$, $b^{(3)}$, $v_\a^{(5)}$, but there is no good reason to reproduce the lengthy expressions for the coefficients here. The crucial property of this solution of the BPS equations is that the coefficients $h^{(1)}$ and $v_\a^{(1)}$ are arbitrary integration constants, while all higher order terms are uniquely determined in terms of these. Notice that the undetermined coefficients $v_\a^{(1)}$ correspond to the VEVs of the dimension one operators, namely  
\be
\cx_-^\a=-\frac{L}{2}v_\a^{(1)},
\ee 
and hence, provided we find a way to fix the integration constant $h^{(1)}$, this solution is the desired BPS swampland that we should use to evaluate the effective action \eqref{extremal-eff-pot-explicit}.

\paragraph{Series resummation and the Cacciatori-Klemm solution} Remarkably, setting the arbitrary integration constants in the solution \eqref{magnetic-uv-exp} to $h^{(1)}=0$ and   
\bal\label{extremal-vs}
v_1^{(1)}=&\;\pm\frac{(k-\frak n_1-\frak n_2)(k-\frak n_1-\frak n_3)}{\sqrt{(k-\frak n_1-\frak n_2)(k-\frak n_1-\frak n_3)(k-\frak n_2-\frak n_3)}},\NO\\
v_2^{(1)}=&\;\pm\frac{(k-\frak n_1-\frak n_2)(k-\frak n_2-\frak n_3)}{\sqrt{(k-\frak n_1-\frak n_2)(k-\frak n_1-\frak n_3)(k-\frak n_2-\frak n_3)}},\NO\\
v_3^{(1)}=&\;\pm\frac{(k-\frak n_1-\frak n_3)(k-\frak n_2-\frak n_3)}{\sqrt{(k-\frak n_1-\frak n_2)(k-\frak n_1-\frak n_3)(k-\frak n_2-\frak n_3)}},
\eal
where the signs are correlated, the expansions for $h(\bar r)$ and $b(\bar r)$ truncate, while those for $v_\a(\bar r)$ can be resummed. The result is the Cacciatori-Klemm solution \cite{Cacciatori:2009iz,Hristov:2010ri}
\bal\label{CK-sol}
&h(\bar r)=\sqrt{\prod_\L\(\a_\L+\frac{\b_\L}{\bar r}\)},\qquad g(\bar r)=\(\frac{\bar r}{L}+\frac{c\;L}{\bar r}\)^2,\\
&v_1=\sqrt{\frac{(\a_2+\b_2/\bar r)(\a_3+\b_3/\bar r)}{(\a_0+\b_0/\bar r)(\a_1+\b_1/\bar r)}},\quad v_2=\sqrt{\frac{(\a_1+\b_1/\bar r)(\a_3+\b_3/\bar r)}{(\a_0+\b_0/\bar r)(\a_2+\b_2/\bar r)}},\quad v_3=\sqrt{\frac{(\a_2+\b_2/\bar r)(\a_1+\b_1/\bar r)}{(\a_0+\b_0/\bar r)(\a_3+\b_3/\bar r)}},\NO
\eal 
where the real constants $\a_\L$, $\b_\L$ and $c$ satisfy the constraints \cite{Cacciatori:2009iz,Hristov:2010ri,Benini:2015eyy}
\be\label{CK-constraints}
\a_\L=1,\qquad \sum_\L\b_\L=0,\qquad \frak n^\L=c+\frac{\b_\L^2}{L^2}.
\ee 

\paragraph{The effective action and its extremization} We now show that the Cacciatori-Klemm solution is obtained by evaluating the effective action \eqref{extremal-eff-pot-explicit} on the solution \eqref{magnetic-uv-exp} with $h^{(1)}=0$, and extremizing with respect to the VEVs $v_\a^{(1)}$. Since we do not know the swampland solution in cosed form, we cannot explicitly obtain the effective action as a function of the VEVs $v^{(1)}_\a$. However, from the expression \eqref{extremal-eff-pot-explicit} follows that the effective action depends on the VEVs $v^{(1)}_\a$ only through the value of the vector multiplet scalars on the horizon, namely  
\be
\left.z^\a\right|_h=z_h^\a(v_\b^{(1)}).
\ee
From the counterterms \eqref{susy-ct} and the definition \eqref{scalar-source-N} of the sources of the dimension one scalar operators follows that on supersymmetric vacua ($\frak m^\L=0$ because we consider purely magnetic solutions here) 
\be
\frac{\d \G_{\rm BPS}[\cx_{-}^\a;\frak m^\L=0,\frak n^\L]}{\d\cx^\a_-}=-J_\a^+=0,
\ee
and so such vacua correspond to the extrema of the effective action. Moreover, provided 
\be
\det\(\frac{\pa z_h^\b}{\pa v_\a^{(1)}}\)\neq 0,
\ee
which we will assume, the chain rule 
\be  
\frac{\pa\G_{\rm BPS}}{\pa v_\a^{(1)}}=\frac{\pa z_h^\b}{\pa v_\a^{(1)}}\frac{\pa\G_{\rm BPS}}{\pa z_h^\b},
\ee
implies that the extrema of the effective action as function the VEVs $v^{(1)}_\a$ correspond to its extrema as function of the values $z_h^\a$ of the scalars on the horizon. This result provides a holographic interpretation of the attractor mechanism, as the extremization of the quantum effective action in the dual theory with respect to the VEVs of the dimension one operators. 

For the purely magnetic black holes we can verify explicitly that extremizing the effective action with respect to the scalars on the horizon, or equivalently the scalar VEVs, reproduces the values $z_h^\a$ on the horizon obtained from the Cacciatori-Klemm solution. From the expression \eqref{extremal-eff-pot-explicit} for the effective action follows that as a function of the values $z_h^\a$ of the scalars on the horizon it takes the form 
\bal
\G_{\rm BPS}\propto &\; (1+z^1_hz^2_hz^3_h)^{-1}\Big(p^0(1-z^1_h)(1-z^2_h)(1-z^3_h)+p^1(1-z^1_h)(1+z^2_h)(1+z^3_h)\NO\\
&+p^2(1+z^1_h)(1-z^2_h)(1+z^3_h)+p^3(1+z^1_h)(1+z^2_h)(1-z^3_h)\Big).
\eal
The extrema of this function are   
\be\label{z-extremal}
z^1_*=\frac{1-\sqrt{\frac{x^2x^3}{x^0x^1}}}{1+\sqrt{\frac{x^2x^3}{x^0x^1}}},\qquad z^2_*=\frac{1-\sqrt{\frac{x^1x^3}{x^0x^2}}}{1+\sqrt{\frac{x^1x^3}{x^0x^2}}},\qquad z^3_*=\frac{1-\sqrt{\frac{x^2x^1}{x^0x^3}}}{1+\sqrt{\frac{x^2x^1}{x^0x^3}}},
\ee
where
\bal
x^0=&\;1+\frac{4(\frak n_0-k/2)^2+1-\frak n_0^2-\frak n_1^2-\frak n_2^2-\frak n_3^2}{2\sqrt{(1-\frak n_0\frak n_1-\frak n_0\frak n_2-\frak n_0\frak n_3-\frak n_1 \frak n_2-\frak n_1\frak n_3-\frak n_2\frak n_3)^2-4\frak n_0\frak n_1\frak n_2\frak n_3}},\NO\\
x^1=&\;1+\frac{4(\frak n_1-k/2)^2+1-\frak n_0^2-\frak n_1^2-\frak n_2^2-\frak n_3^2}{2\sqrt{(1-\frak n_0\frak n_1-\frak n_0\frak n_2-\frak n_0\frak n_3-\frak n_1 \frak n_2-\frak n_1\frak n_3-\frak n_2\frak n_3)^2-4\frak n_0\frak n_1\frak n_2\frak n_3}},\NO\\
x^2=&\;1+\frac{4(\frak n_2-k/2)^2+1-\frak n_0^2-\frak n_1^2-\frak n_2^2-\frak n_3^2}{2\sqrt{(1-\frak n_0\frak n_1-\frak n_0\frak n_2-\frak n_0\frak n_3-\frak n_1 \frak n_2-\frak n_1\frak n_3-\frak n_2\frak n_3)^2-4\frak n_0\frak n_1\frak n_2\frak n_3}},\NO\\
x^3=&\;1+\frac{4(\frak n_3-k/2)^2+1-\frak n_0^2-\frak n_1^2-\frak n_2^2-\frak n_3^2}{2\sqrt{(1-\frak n_0\frak n_1-\frak n_0\frak n_2-\frak n_0\frak n_3-\frak n_1 \frak n_2-\frak n_1\frak n_3-\frak n_2\frak n_3)^2-4\frak n_0\frak n_1\frak n_2\frak n_3}}.
\eal

Using the solution of the conditions \eqref{CK-constraints}, namely
\bal
\b_\L=&\;\pm\frac{L}{4}\(\frac{4(\frak n_\L-\frac k2)^2+1-(\frak n_0^2+\frak n_1^2+\frak n_2^2+\frak n_3^2)}{\sqrt{(k-\frak n_1-\frak n_2)(k-\frak n_1-\frak n_3)(k-\frak n_2-\frak n_3)}}\),\quad k=\pm1,\NO\\
c=&\;\frac{k}{2}-\frac{\b_1^2+\b_2^2+\b_3^2+\b_1\b_2+\b_2+\b_3+\b_1\b_3}{2L^2},
\eal
it is straightforward to verify that the values \eqref{z-extremal} of the scalars on the horizon are exactly those obtained from the solution \eqref{CK-sol}, and hence, the corresponding value of the effective action coincides with the black hole entropy. Moreover, since we know this solution in closed form, we determine that the scalar VEVs that extremize the effective potential are given by \eqref{extremal-vs}, or equivalently 
\bal\label{extremal-VEVs-update}
\<\co^1_{\D=1}\>=&\;\pm\frac{L}{2}\(\frac{(k-\frak n_1-\frak n_2)(k-\frak n_1-\frak n_3)}{\sqrt{(k-\frak n_1-\frak n_2)(k-\frak n_1-\frak n_3)(k-\frak n_2-\frak n_3)}}\),\NO\\
\<\co^2_{\D=1}\>=&\;\pm\frac{L}{2}\(\frac{(k-\frak n_1-\frak n_2)(k-\frak n_2-\frak n_3)}{\sqrt{(k-\frak n_1-\frak n_2)(k-\frak n_1-\frak n_3)(k-\frak n_2-\frak n_3)}}\),\NO\\
\<\co^3_{\D=1}\>=&\;\pm\frac{L}{2}\(\frac{(k-\frak n_1-\frak n_3)(k-\frak n_2-\frak n_3)}{\sqrt{(k-\frak n_1-\frak n_2)(k-\frak n_1-\frak n_3)(k-\frak n_2-\frak n_3)}}\),
\eal
with the overall signs correlated.

The above analysis demonstrates that the entropy functional for purely magnetic BPS solutions of the $U(1)^4$ theory should be identified with the quantum effective action for the dimension one scalar operators in the twisted ABJM model at Chern-Simons level one. In particular, the purely magnetic black holes correspond to zero chemical potentials for the currents $\cj_\L^i$. The value of the quantum effective action on its extremum coincides with the extremal black hole entropy, as well as the Witten index in the twisted ABJM model. In order to turn on the chemical potentials in the supersymmetric index discussed in \cite{Benini:2015noa,Benini:2015eyy,Benini:2016rke} it is mandatory to consider dyonic black holes in the bulk -- it is not possible for the chemical potentials to be non-zero for electrically neutral black holes. With non-zero electric chemical potentials $\m^\L$, the supersymmetric index coincides with the quantum effective action for the dimension one operators in the twisted ABJM model at Chern-Simons level one, {\em deformed} by the supersymmetric fermion masses $\frak m^\L=\m^\L$. This effective action, given by \eqref{extremal-eff-pot-explicit}, coincides with the black hole entropy at the extremal value of the scalar VEVs for the dimension one operators.

\subsection{Dyonic BPS black holes}

As we have just argued, the supersymmetric index discussed in \cite{Benini:2015noa,Benini:2015eyy,Benini:2016rke}, with arbitrary fugacities, should be matched to the quantum effective action \eqref{extremal-eff-pot-explicit}, evaluated on the BPS swampland with non-zero chemical potentials. The corresponding dyonic BPS solution can be sought in the form of a UV expansion, analogous to \eqref{magnetic-uv-exp} for the purely magnetic solutions. As we discussed at the beginning of this section, at fixed chemical potentials this solution depends on $n_V$ arbitrary VEVs for the dimension one scalar operators. 

The identity \eqref{extremum-Y} implies that the BPS effective action is extremized with respect to the sources $\cy_-^\a$ of the dimension two operators, with the extremal values of these sources related to the electric chemical potentials as in \eqref{susy-sources}. It follows that further extremizing the effective action with respect to the VEVs $\cx_-^\a$ of the dimension one operators amounts to extremizing the effective action with respect to the complex modes $z_-^\a=\cx^\a_- +i \cy^\a_-$, which appear at leading order in the Fefferman-Graham expansion of the scalars $z^\a$, given in \eqref{FG-FP}. The same argument as for the magnetic black holes and real scalars above then implies that extremizing the effective action with respect to $z_-^\a=\cx^\a_- +i \cy^\a_-$ is equivalent to extremizing it with respect to the {\em complex} values $z_h^\a$ of the scalars on the horizon. The latter corresponds to the attractor mechanism \cite{Benini:2016rke} and this result, therefore, extends our holographic interpretation of the attractor mechanism as the extremization of the quantum effective action to general dyonic BPS solutions. The BPS black solutions corresponding to the extremum of the effective action in the dyonic case are those found in \cite{Halmagyi:2014qza} and the corresponding entropy functional was discussed in \cite{Goulart:2015lwd}. These solutions can be used to obtain the values of the VEVs of the dimension one operators at the extremum of the effective action in terms of the magnetic and electric charges, but we will not compute these VEVs explicitly here.

\section{Concluding remarks}
\label{conclusion}
\setcounter{equation}{0}

One of our main results is a clarification of the holographic renormalization paradigm for asymptotically AdS$_4$ black holes in ${\cal N}=2$ gauged  supergravity. Along these lines and with the hope of providing a purely field theoretic interpretation for some of the assumptions made in the comparison with the microscopic entropy via topologically twisted index computations, we have found a boundary interpretation for the attractor mechanism.  

Our conceptual home for the attractor mechanism in asymptotically AdS spacetimes that are solutions of ${\cal N}=2$ gauged  supergravity shows that it is equivalent, on the field theory side, to extremizing the quantum effective action with respect to certain VEVs. Our formulation of the mechanism retains some features of the original formulation in asymptotically flat spacetimes but exploits the inner workings of the AdS/CFT dictionary. For example, instead of extremizing with respect to moduli, as in the asymptotically flat case, we extremize with respect to VEVs in the asymptotically AdS case.  Rather than extremizing the central charge that appears in the original formulation of attractor mechanism, we extremize the quantum effective action as follows from the entropy formula formalism. 

We have resolved some conceptually challenging issues in the path toward the identification of the topologically twisted index and black hole entropy. In particular, we have clarified the nature of: (i) the field theoretic need for extremization and its connection with the attractor mechanism, (ii) the proper identification of scalar VEVs and the precise relations to the chemical potentials.

There are a number of open problems that would be interesting to tackle using the results we have obtained in this work. An obvious but technically challenging problem is to repeat our analysis for rotating asymptotically AdS$_4$ black holes and, more importantly, asymptotically AdS$_5$ ones. For rotating black holes it is much more difficult to obtain first order equations and to derive general expressions for the on-shell action, but AdS$_5$ black holes introduce additional complications of a completely different nature. In particular, supersymmetry on four dimensional curved backgrounds is generically anomalous \cite{Papadimitriou:2017kzw,An:2017ihs}\footnote{This anomaly was implicitly present in the analysis of \cite{Genolini:2016ecx} as well, but was not recognized as such.}, which leads to anomalous contributions in the BPS relations among the conserved charges of supersymmetric AdS$_5$ black holes.    

Moreover, having clarified the connection with the entropy formula, it would be quite interesting to extend our findings to include an interpretation of the quantum entropy formula \cite{Sen:2008vm}. Indeed, after a preliminary discussion in \cite{Liu:2017vll} and \cite{Jeon:2017aif} focusing on the near horizon degrees of freedom,  some quantum corrections to the black hole entropy have been matched using an approach that focuses on the asymptotic degrees of freedom \cite{Liu:2017vbl}.

\section*{Acknowledgments}

We thank Francesco Benini, Davide Cassani, Alejandra Castro, Mirjam Cveti\v c, Alessandra Gnecchi, Rajesh Gupta, Nick Halmagyi, Seyed Morteza Hosseini, Kiril Hristov, Finn Larsen, Dario Martelli, Sameer Murthy, Silviu Pufu, Valentin Reys, Chiara Toldo, Stefan Vandoren and Alberto Zaffaroni for helpful discussions.  The work of ACB is supported by  the ERC Consolidator Grant N. 681908, “Quantum black holes: A macroscopic window into the microstructure of gravity”. LPZ and VR are  partially supported by the US Department of Energy under Grant No. \ DE-SC0017808 -- {\it Topics in the AdS/CFT Correspondence: Precision tests with Wilson loops, quantum black holes and dualities}. The work of UK is supported in part by NSF through grant PHY-1620039.

\appendix

\renewcommand{\theequation}{\Alph{section}.\arabic{equation}}

\setcounter{section}{0}

\section*{Appendix}
\setcounter{section}{0}

\section{Parameterizations of the 4D $\cn=2$ $U(1)^4$ gauged supergravity}
\label{STU-parameterizations}

In this appendix we summarize some of the common choices for the symplectic sections of the $U(1)^4$ $\cn=2$ gauged supergravity, and we provide the explicit relations for the physical scalars between different parameterizations. 
As we pointed out in section \ref{N=2}, the K\"ahler potential and the holomorphic superpotential transform non-trivially under reparameterizations of the symplectic sections and, therefore, their expressions in two different parameterizations of the symplectic sections should not be identified.  Moreover, as is evident from section \ref{holren}, not all choices of holomorphic sections are compatible with a particular choice of boundary conditions on the scalars. As a result, only certain choices for the symplectic sections are compatible with supersymmetric boundary conditions and/or holography.

\subsection{Cveti\v c et al. gauge} 
\label{Cvetic-par}

The choice of symplectic sections that leads to the original parameterization of the STU model in \cite{Cvetic:1999xp} is summarized in section 8 of \cite{Hristov:2010ri}, and for the special case of real $X^\L$ also in appendix A.1 of \cite{Benini:2015eyy}. The relevant parameterization is 
\be\label{Cvetic-par-ratios}
\frac{X^1}{X^0}\equiv \t_2\t_3,\qquad \frac{X^2}{X^0}\equiv \t_1\t_3,\qquad \frac{X^3}{X^0}\equiv \t_1\t_2,
\ee
together with the gauge fixing condition
\be\label{Cvetic-par-gauge}\boxed{
X^0X^1X^2X^3=1,}
\ee
or equivalently
\be
X^0=\frac{1}{\sqrt{\t_1\t_2\t_3}}.
\ee
With this choice of symplectic sections, the prepotential, the K\"ahler potential, the K\"ahler metric, the holomorphic superpotential and the scalar potential defined in section \ref{N=2} take respectively the form
\bal\label{Cvetic-par-results}
F=&\;-2i,\NO\\
\ck=&\;-\log\Big(\frac{(\t_1+\bar\t_1)(\t_2+\bar\t_2)(\t_3+\bar\t_3)}{|\t_1\t_2\t_3|}\Big),\NO\\
\ck_{\a\bar\b}=&\;\frac{\d_{\a\bar\b}}{(\t_\a+\bar\t_{\bar\b})^2},\NO\\
W=&\;\x\(\frac{1+\t_1\t_2+\t_1\t_3+\t_2\t_3}{\sqrt{\t_1\t_2\t_3}}\),\NO\\
\cv=&\;-\frac{2}{L^2}\(\frac{1+|\t_1|^2}{\t_1+\bar\t_1}+\frac{1+|\t_2|^2}{\t_2+\bar\t_2}+\frac{1+|\t_3|^2}{\t_3+\bar\t_3}\).
\eal
Setting further 
\be\label{tau-par}\boxed{
	\t_\a=e^{-\vf_\a}+i\c_\a,}
\ee
it is straightforward to show that the $\cn=2$ action \eqref{action} reduces to the STU model action given in \cite{Cvetic:1999xp}. Notice that this parameterization is particularly convenient in the case of real $\t^\a$ because the K\"ahler potential becomes a constant. This in turn implies that the holomorphic superpotential coincides with the real superpotential \eqref{real-sup}, given in eq.~(3.1) or \cite{Papadimitriou:2006dr} (see also eq.~(2.15) in \cite{Duff:1999gh}). 

A very important property of the parameterization of the STU model in terms of the scalars $\vf_\a$ and $\c_\a$ is that it is compatible with the holographic dictionary, since these scalars have the correct Fefferman-Graham expansions for fields dual to dimension one or dimension two operators. Namely,
\bal\label{FG-STU}
\vf_\a(r,x)=&\;\vf^-_{\a}(x)e^{-r/L}+\vf^+_{\a}(x)e^{-2r/L}+\cdots,\NO\\
\c_\a(r,x)=&\;\c^-_{\a}(x)e^{-r/L}+\c^+_{\a}(x)e^{-2r/L}+\cdots,
\eal
in the Fefferman-Graham coordinates defined by the metric \eqref{ADM-metric} and the gauge-fixing conditions \eqref{FG-gauge}. This parameterization is also compatible with supersymmetry, which requires that Neumann boundary conditions be imposed on the dilatons $\vf_\a$ and Dirichlet on the axions $\c_\a$ (or vice versa) \cite{Breitenlohner:1982bm}.

\subsection{Cacciatori-Klemm gauge}
\label{CK-par}

A related parameterization of the symplectic sections is used in section 3.2 of \cite{Cacciatori:2009iz}, where the 
ratios 
\be
\frac{X^1}{X^0}\equiv \t_2\t_3,\qquad \frac{X^2}{X^0}\equiv \t_1\t_3,\qquad \frac{X^3}{X^0}\equiv \t_1\t_2,
\ee
are parameterized exactly as in \eqref{Cvetic-par-ratios}, but the gauge condition \eqref{Cvetic-par-gauge} is replaced with 
\be\boxed{
X^0=1.}
\ee
This choice leads to 
\bal\label{CK-par-results}
F=&\;-2i\t_1\t_2\t_3,\NO\\
\ck=&\;-\log\Big((\t_1+\bar\t_1)(\t_2+\bar\t_2)(\t_3+\bar\t_3)\Big),\NO\\
\ck_{\a\bar\b}=&\;\frac{\d_{\a\bar\b}}{(\t_\a+\bar\t_{\bar\b})^2},\NO\\
W=&\;\x(1+\t_2\t_3+\t_1\t_3+\t_1\t_2),\NO\\
\cv=&\;-\frac{2}{L^2}\(\frac{1+|\t_1|^2}{\t_1+\bar\t_1}+\frac{1+|\t_2|^2}{\t_2+\bar\t_2}+\frac{1+|\t_3|^2}{\t_3+\bar\t_3}\).
\eal
Notice that the K\"ahler metric and the scalar potential are identical to those in \eqref{Cvetic-par-results}, and so using the identification \eqref{tau-par} one again obtains the STU model Lagrangian in the form given in \cite{Cvetic:1999xp}. However, the K\"ahler potential and the holomorphic superpotential are not the same, which is expected since neither of these quantities is invariant under reparameterizations of the symplectic sections. Despite the different gauge fixing condition for the symplectic sections, the Cacciatori-Klemm parameterization leads to the same physical scalars as those in Cveti\v c et al., and so it is also compatible with both the holographic dictionary and supersymmetry.

\subsection{Pufu-Freedman gauge}

Another choice for the symplectic sections that is compatible with both holography and supersymmetry is the one implicitly used in \cite{Breitenlohner:1982bm,Freedman:2013ryh}. In that parameterization the sections are given by
\bal\label{PF-sections}
X^0=&\;(1+z^1)(1+z^2)(1+z^3),\NO\\
X^1=&\;(1+z^1)(1-z^2)(1-z^3),\NO\\
X^2=&\;(1-z^1)(1+z^2)(1-z^3),\NO\\
X^3=&\;(1-z^1)(1-z^2)(1+z^3),
\eal
from which we obtain  
\bal\label{FP-par-results}
F=&\;-2i(1-(z^1)^2)(1-(z^2)^2)(1-(z^3)^2),\NO\\
\ck=&\;-\log\Big(8(1-|z^1|^2)(1-|z^2|^2)(1-|z^3|^2)\Big),\NO\\
\ck_{\a\bar\b}=&\;\frac{\d_{\a\bar\b}}{(1-|z^\a|^2)^2},\NO\\
W=&\;4\x(1+z^1z^2z^3),\NO\\
\cv=&\;\frac{2}{L^2}\Big(3-2\sum_{\a=1}^3\frac{1}{1-|z^\a|^2}\Big).
\eal
The physical scalars $z^\a$ in this parameterization are related to the variables $\t^\a$ in the Cveti\v c et al. and Cacciatori-Klemm gauges as 
\be\label{scalar-map}\boxed{
z^1=\frac{1-\t_1}{1+\t_1},\qquad z^2=\frac{1-\t_2}{1+\t_2},\qquad z^3=\frac{1-\t_3}{1+\t_3}.}
\ee
These scalars also admit the correct Fefferman-Graham expansions for fields dual to dimension one or two operators, namely  
\be\label{FG-FP}
z^\a(r,x)=z_-^{\a}(x)e^{-r/L}+z_+^{\a}(x)e^{-2r/L}+\cdots.
\ee
The relations \eqref{scalar-map} then imply that the modes $z_-^{\a}(x)$ and $z_+^{\a}(x)$ can be expressed in terms of the modes of the Fefferman-Graham expansions \eqref{FG-STU} as
\begin{align}\label{mode-map}
\boxed{
\begin{aligned}
z_-^{\a}(x)=&\;\frac12(\vf^-_{\a}(x)-i \c^-_{\a}(x)),\\ z_+^{\a}(x)=&\;\frac12\Big[\Big(\vf^+_{\a}(x)-\frac12(\c^-_{\a}(x))^2\Big)-i \Big(\c^+_{\a}(x)+\vf^-_{\a}(x)\c^-_{\a}(x)\Big)\Big],
\end{aligned}
}
\end{align}
where no summation over the index $\a$ is implied. Hence, the boundary conditions for $\vf_\a$ and $\c_\a$ map respectively to the real and imaginary parts of $z^\a$.

\subsection{Hristov-Vandoren gauge}

As a final example of a choice of symplectic sections for the STU model we should discuss the parameterization used in \cite{Hristov:2010ri,Benini:2015eyy}, namely
\bal
X^0=&\;\frac{1}{1+\tilde z^1+\tilde z^2+\tilde z^3},\NO\\ 
X^1=&\;\frac{\tilde z^1}{1+\tilde z^1+\tilde z^2+\tilde z^3},\NO\\
X^2=&\;\frac{\tilde z^2}{1+\tilde z^1+\tilde z^2+\tilde z^3},\NO\\
X^3=&\;\frac{\tilde z^3}{1+\tilde z^1+\tilde z^2+\tilde z^3},
\eal
together with the gauge fixing condition (see discussion around eq.~(4.16) in \cite{Hristov:2010ri} and eq.~(C.4) in \cite{Benini:2015eyy})
\be
X^0+X^1+X^2+X^3=1,
\ee
as well as the reality condition
\be
\Im X^\L=0.
\ee

In order to compute the various $\cn=2$ supergravity quantities one needs to start with general complex $\tilde z^\a$ and impose the reality condition  only at the end. This procedure gives
\bal\label{HV-par-results}
F=&\;\frac{-2i\sqrt{\tilde z^1\tilde z^2\tilde z^3}}{(1+\tilde z^1+\tilde z^2+\tilde z^3)^2},\NO\\
\ck=&\;-\log\(\frac{8\sqrt{\tilde z^1\tilde z^2\tilde z^3}}{(1+\tilde z^1+\tilde z^2+\tilde z^3)^2}\),\quad \tilde z^1,\tilde z^2,\tilde z^3\in\bb R,	\NO\\
\ck_{\a\bar\b}=&\;\frac{1}{16}\left(\begin{matrix}
		3/(\tilde z^1)^2 & -1/\tilde z^1\tilde z^2 & -1/\tilde z^1\tilde z^3\\
		-1/\tilde z^1\tilde z^2 & 3/(\tilde z^2)^2 & -1/\tilde z^2\tilde z^3\\
		-1/\tilde z^1\tilde z^3 & -1/\tilde z^3\tilde z^2 & 3/(\tilde z^3)^2
\end{matrix}\right),\quad \tilde z^1,\tilde z^2,\tilde z^3\in\bb R,\NO\\
W=&\;\x,\NO\\
\cv=&\;-\frac{1}{L^2}\(\frac{\tilde z^1+\tilde z^2+\tilde z^3+\tilde z^1\tilde z^2+\tilde z^1\tilde z^3+\tilde z^2\tilde z^2}{\sqrt{\tilde z^1\tilde z^2\tilde z^3}}\),\quad \tilde z^1,\tilde z^2,\tilde z^3\in\bb R,		
\eal
where we have given the K\"ahler potential, the K\"ahler metric and the scalar potential only for real $\tilde z^\a$ since the expressions with complex scalars are far too lengthy. The expressions for the prepotential and the holomorphic superpotential hold for complex scalars.  

The scalars $\tilde z^\a$ are related to the variables $\t^\a$ in the Cveti\v c et al. and Cacciatori-Klemm parameterizations as
\be\boxed{
\tilde z^1=\t_2\t_3,\qquad \tilde z^2=\t_1\t_3,\qquad \tilde z^3=\t_1\t_2.}
\ee
Taking $\t^\a$ to be real and inserting these expressions in the scalar potential in \eqref{HV-par-results} one easily sees that it coincides with the scalar potential in \eqref{Cvetic-par-results} or \eqref{CK-par-results}. It follows that the parameterization used in \cite{Hristov:2010ri,Benini:2015eyy} agrees with all other parameterizations of the STU model discussed above, but only provided the scalars are {\em real}. It is in fact a very convenient parameterization for obtaining the purely magnetic solutions discussed in \cite{Hristov:2010ri,Benini:2015eyy}, but it is not suitable for supersymmetric dyonic solutions that are necessarily supported by complex scalars.

\section{Radial Hamiltonian formalism}
\label{ham}

In order to formulate the supergravity theory described by the action \eqref{action} in radial Hamiltonian language we parameterize the bulk metric in terms of the lapse function $N$, the shift function $N_i$ and the induced metric $\g_{ij}$ on the radial slices, namely
\be\label{ADM-metric}
ds^2=(N^2+N_iN^i)dr^2+2N_idrdx^i+\g_{ij}dx^idx^j.
\ee
Similarly, the Abelian gauge fields are decomposed in radial and transverse components as 
\be
A^\L=\a^\L dr+A^\L_idx^i.
\ee

Using the decomposition \eqref{ADM-metric} of the metric the bulk Ricci scalar becomes
\be\label{Ricci}
R[g]=R[\g]+K^2-K_{ij}K^{ij}+\nabla_{\mu}\left(-2Kn^{\mu}+2n^{\nu}\nabla_{\nu}n^{\mu}\right),
\ee
where
\be
K_{ij}=\frac{1}{2N}\left(\dot{\g}_{ij}-D_iN_j-D_jN_i\right),
\ee
is the extrinsic curvature and $n^{\mu}=(1/N,-N^i/N)$ is the unit outward normal vector of the radial slices. As in the main text, a dot $\dot{}$ denotes a derivative with respect to the radial coordinate $r$, and $D_i$ is the covariant derivative with respect to the induced metric $\g_{ij}$. Using these expressions, together with the identities 
\be
\sqrt{-g}=N\sqrt{-\g},\qquad g^{\mu\nu}=\begin{pmatrix}\frac{1}{N^2}&-\frac{N^i}{N^2}\\-\frac{N^i}{N^2}&\g^{ij}+\frac{N^iN^j}{N^2}\end{pmatrix},
\ee
the action (\ref{action}) can be written in the form $S=\int dr L$, where  
the radial Lagrangian $L$ is
\begin{align}\label{radial-lagrangian}
L=&\;\frac{1}{2\k^2}\int d^3x N\sqrt{-\g}\Big\{R[\g]+K^2-K_{ij}K^{ij}-\frac{1}{N^2}\cg_{\a\bar\b}\big(\dot{z}^\a-N^i\pa_i z^\a\big)\big(\dot{\bar z}^{\bar\b}-N^i\pa_i\bar z^{\bar \b}\big)\NO\\
&-\frac{4}{N^2}\ci_{\L\S}\g^{ij}(\dot{A}_i^\L-\pa_i\a^\L-N^kF^\L_{ki})(\dot{A}^\S_j-\pa_j\a^\S-N^lF^\S_{lj})-\frac{4}{N}\car_{\L\S}\e^{ijk}(\dot{A}^\L_i-\pa_i\a^\L)F^\S_{jk}\NO\\
&-\cg_{\a\bar\b}\g^{ij}\pa_iz^\a\pa_j\bar z^{\bar\b}-2\ci_{\L\S}F^\L_{ij}F^{\S ij}-\cv\Big\}.
\end{align}

The canonical momenta following from this Lagrangian take the form
\begin{subequations}\label{momenta}
	\begin{align}
	\pi^{ij}=&\frac{\d L}{\d\dot{\g}_{ij}}=\frac{1}{2\k^2}\sqrt{-\g}\big(K\g^{ij}-K^{ij}\big), \\
	\pi_{\a}=&\frac{\d L}{\d\dot{z}^\a}=-\frac{1}{2\k^2}\frac{\sqrt{-\g}}{N}\cg_{\a\bar\b}\big(\dot{\bar z}^{\bar \b}-N^i\pa_i\bar z^{\bar \b}\big), \\
	\pi_{\bar\b}=&\frac{\d L}{\d\dot{\bar z}^{\bar \b}}=-\frac{1}{2\k^2}\frac{\sqrt{-\g}}{N}\cg_{\a\bar\b}\big(\dot{z}^\a-N^i\pa_iz^\a\big), \\
	\pi^i_\L=&\frac{\d L}{\d\dot{A}_i^\L}=-\frac{4}{\k^2}\frac{\sqrt{-\g}}{N}\ci_{\L\S}\big(\g^{ij}(\dot{A}^\S_j-\pa_j\a^\S)-N_jF^{\S ji}\big)-\frac{2}{\k^2}\sqrt{-\g}\car_{\L\S}\e^{ijk}F^\S_{jk}.
	\end{align}
\end{subequations}
Notice that the canonical momenta conjugate to the variables $N$, $N_i$ and $\a^\L$ vanish identically and, hence, these fields are non-dynamical. Given the canonical momenta \eqref{momenta}, a short calculation determines the Hamiltonian, namely
\be\label{radial-hamiltonian}
H=\int d^3x\Big(\pi^{ij}\dot{\g}_{ij}+\pi_{\a}\dot{z}^\a+\pi_{\bar\b}\dot{\bar z}^{\bar\b}+\pi_\L^i\dot{A}^\L_i\Big)-L=\int d^3x\left(N\mathcal{H}+N_i\mathcal{H}^i+\a^\L\cf_\L\right),
\ee
where
\begin{subequations}\label{constraints}
	\begin{align}
	\mathcal{H}=&-\frac{\k^2}{\sqrt{-\g}}\left(2\left(\g_{ik}\g_{jl}-\frac12\g_{ij}\g_{kl}\right)\pi^{ij}\pi^{kl}\right.\NO\\
	&\left.+\frac{1}{8}\ci^{\L\S}\left(\pi_{\L i}+\frac{2}{\k^2}\sqrt{-\g}\;\car_{\L K}\e_i{}^{kl}F^K_{kl}\right)\left(\pi^i_\S+\frac{2}{\k^2}\sqrt{-\g}\;\car_{\S M}\e^{ipq}F^M_{pq}\right)+2\cg^{\a\bar\b}\p_\a\p_{\bar\b}\right) \NO \\
	&+\frac{\sqrt{-\g}}{2\k^2}\left(-R[\g]+2\ci_{\L\S}F^\L_{ij}F^{\S ij}+\cg_{\a\bar\b}\pa_i z^\a\pa^i\bar z^{\bar\b}+\cv\right), \\
	\mathcal{H}^i=&-2D_j\pi^{ij}+F^{\L ij}\left(\pi_{\L j}+\frac{2}{\k^2}\sqrt{-\g}\;\car_{\L\S}\e_j{}^{kl}F^\S_{kl}\right)+\pi_{\a}\pa^i z^\a+\p_{\bar\b}\pa^i\bar z^{\bar\b}, \\
	\cf_\L=&-D_i\pi_\L^i.
	\end{align}
\end{subequations}
Since the canonical momenta conjugate to the fields $N$, $N_i$ and $\a^\L$ vanish identically, Hamilton's equations for these fields impose the first class constraints 
\be\label{constraints=0}
\ch=\ch^i=\cf_\L=0,
\ee 
which reflect the diffeomorphism and gauge invariance of the bulk theory. It follows that the Hamiltonian \eqref{radial-hamiltonian} vanishes identically on-shell.  

Finally, HJ theory allows us to express the canonical momenta as gradients of the so called Hamilton's principal function $\cs[\g,A^\L,z^\a,\bar z^{\bar\b}]$, i.e.
\be\label{HJ-momenta}
\pi^{ij}=\frac{\d \cs}{\d\g_{ij}},\quad \pi^i_\L=\frac{\d \cs}{\d A_i^\L},\quad \pi_\a=\frac{\d \cs}{\d z^\a},\quad \pi_{\bar\b}=\frac{\d \cs}{\d\bar z^{\bar\b}}. 
\ee
Inserting these expressions for the momenta in the constraints \eqref{constraints} leads to a set of functional partial differential equations, the HJ equations, for the functional $\cs[\g,A^\L,z^\a,\bar z^{\bar\b}]$. Given a solution of the HJ equations, equating the expressions \eqref{HJ-momenta} and \eqref{momenta} for the canonical momenta leads to a set of first order flow equations for the fields $\g_{ij}(r,x)$, $A_i^\L(r,x)$, $z^\a(r,x)$, $\bar z^{\bar\b}(r,x)$. In the radial (or Fefferman-Graham) gauge  
\be
\label{FG-gauge}
N=1,\qquad N_i=0,\qquad  \a^\L=0,
\ee
these first order equations take the form 
\begin{subequations}\label{flow}
	\begin{align}
	\dot{\g}_{ij}=&-\frac{4\k^2}{\sqrt{-\g}}\left(\g_{ik}\g_{jl}-\frac12\g_{ij}\g_{kl}\right)\frac{\d \cs}{\d\g_{kl}}, \\
	\dot{z}^\a=&-\frac{2\k^2}{\sqrt{-\g}}\cg^{\a\bar\b}\frac{\d \cs}{\d\bar z^{\bar\b}}, \\
	\dot{\bar z}^{\bar\b}=&-\frac{2\k^2}{\sqrt{-\g}}\cg^{\a\bar\b}\frac{\d \cs}{\d z^\a} ,\\
	\dot{A}^\L_i=&-\frac{\k^2}{4\sqrt{-\g}}\ci^{\L\S}\g_{ij}\frac{\d \cs}{\d A_j^\S}-\frac12\ci^{\L\S}\car_{\S M}\e_i\,^{jk}F^M_{jk}.
	\end{align}
\end{subequations}
As we discuss in section \ref{eff-sup}, these general flow equations lead to first order BPS-like equations for any solution of the form \eqref{Bans}, including non supersymmetric solutions.


\bibliographystyle{jhepcap}
\bibliography{refs}

\end{document}